\documentclass[12pt]{article}

\usepackage{latexsym}
\usepackage{verbatim}
\usepackage{setspace}
\usepackage{color}
\usepackage{physics}
\usepackage{tikz}
\usepackage{mathdots}
\usepackage{yhmath}
\usepackage{cancel}
\usepackage{color}
\usepackage{siunitx}
\usepackage{array}
\usepackage{multirow}
\usepackage{gensymb}
\usepackage{tabularx}
\usepackage{booktabs}
\usetikzlibrary{fadings}
\usetikzlibrary{patterns}
\usetikzlibrary{shadows.blur}
\usetikzlibrary{shapes}
\usepackage{cite}
\usepackage{hyperref}
\usepackage{cleveref}
\usepackage[mathscr]{euscript}
\usepackage{changes}
\usepackage{cite}
\usepackage{xspace}
\usepackage{xparse}
\usepackage{mathtools}
\usepackage{mathrsfs}
\usepackage{caption}
\usepackage{subcaption}
\usepackage{tensor}
\usepackage{amsfonts}
\usepackage{amssymb}
\usepackage{amsmath,amssymb}
\usepackage[utf8]{inputenc}
\usepackage{url}

\newcommand{\sgn}{{\rm sgn}}

\newcommand{\vac}{0}

\newcommand{\bit}{\begin{itemize}}
\newcommand{\eit}{\end{itemize}}
\newcommand{\bd}{\begin{description}}
\newcommand{\ed}{\end{description}}
\newcommand{\bc}{\begin{center}}
\newcommand{\ec}{\end{center}}
\newcommand{\be}{\begin{equation}}
\newcommand{\ee}{\end{equation}}
\newcommand{\bea}{\begin{eqnarray}}
\newcommand{\eea}{\end{eqnarray}}
\newcommand{\bs}{\begin{subequations}}
\newcommand{\es}{\end{subequations}}

\def\bz{\bar{z} }

\newcommand{\wightmanthree}[3]{\mathcal{W}_{#1 #2 #3}}
\newcommand{\spherevolume}[1]{\operatorname{Vol}(S^{#1})}

\newcommand{\Ni}{N_{\Delta_i}}

\numberwithin{equation}{section}

\begin{document}

\begin{titlepage}

\begin{center}
{\Large{{\bf Celestial amplitudes from conformal \vspace{5pt}}}}\\
{\Large{{\bf correlators with bulk-point kinematics}}\vspace{10pt}}
\date{}
\vspace{10pt}\\
Leonardo Pipolo de Gioia$^{*}$ and Ana-Maria Raclariu$^{\dagger}$\\
\vspace{0.8cm}
\small{$^{*}$\textit{ICTP South American Institute for Fundamental Research\\IFT-UNESP, S\~{a}o Paulo, SP Brazil 01440-070} }\\  \small{$^{\dagger}$\textit{King's College London, Strand, London WC2R 2LS, United Kingdom}}
\vspace{30pt}
\begin{abstract}

We show that two- and three-point celestial (C)CFT$_{d-1}$ amplitudes can be directly obtained from correlation functions in a unitary Lorentzian CFT$_d$ on $\mathbb{R}\times S^{d-1}$. The recipe involves a rescaling of the operators, followed by an expansion around a bulk point configuration and a transformation to an  $S^{d-1}$ conformal primary basis. The first two steps project the CFT$_d$ correlators onto distributions on $S^{d-1}$. The final step implements a dimensional reduction yielding CCFT$_{d-1}$ amplitudes that are manifestly vanishing for all in/out configurations and Poincar\'e invariant. The dimensional reduction may be implemented either by evaluating certain time integral transforms around the bulk-point limit, or by analytically continuing the  CFT$_d$ operator dimensions and restricting the operators to $S^{d-1}$ time slices separated by $\pi$ in global time. The latter prescription generates the correct normalization for both two- and three-point functions. On the other hand, the celestial three-point amplitudes obtained via the former prescription are found to only agree after evaluating a residue at an integer linear combination of the CFT$_d$ conformal dimensions. The correct normalization may also be obtained by considering a different integration path in the uplift of the complexified time plane to its universal cover.
\end{abstract}

\end{center}

\end{titlepage}

\newpage
\tableofcontents

\section{Introduction}

Celestial holography proposes a correspondence between quantum theories of gravity in asymptotically flat spacetimes (AFS) and codimension two celestial conformal field theories (CCFT) defined at null infinity. This proposal is motivated by the realization that the asymptotic symmetry structure of 3+1-dimensional AFS leads to the same constraints on bulk scattering amplitudes imposed by low-energy theorems \cite{He:2014laa,Cachazo:2014fwa,Kapec:2014opa,Lysov:2014csa,He:2014cra}. When expressed in a basis of boost eigenstates (also known as a conformal primary basis), S-matrix elements share similarities with correlation functions of primary operators in a 2-dimensional conformal field theory (CFT).
In contrast to the well-established AdS/CFT correspondence, the development of celestial holography has predominantly followed a bottom-up approach, with the challenging task of constructing dual pairs in a top-down manner remaining largely unexplored, except for a few recent proposals primarily focused on the self-dual sector \cite{costello2023top, costello2023burns}.

In the quest for top-down models of celestial holography one is naturally led to explore its relation to the well-studied Anti de Sitter-Conformal Field Theory (AdS/CFT) correspondence. Two distinct lines of inquiry have emerged so far. One aims to better understand the $(d-1)$-dimensional CCFT by regarding it as defined on the common boundary of families of $(d)$-dimensional AdS and dS slices foliating the Milne subregions of $(d+1)$-dimensional Minkowski spacetime and their complement \cite{deBoer:2003vf, Cheung:2016iub, Iacobacci:2022yjo, Sleight:2023ojm, melton2023celestial, iacobacci2024celestial, melton2024soft, melton2024celestial}. The  other relies on examining the flat space limit of $(d+1)$-dimensional AdS spacetimes and their holographic $d$-dimensional CFT duals, which has proven fruitful in understanding the structure of scattering amplitudes (particularly of massive particles) in flat space.

In this paper we further explore the second possibility, motivated by the recent observation that AdS-Witten diagrams in Lorentzian signature reduce to celestial amplitudes of massless particles in a particular limit. Specifically, it was shown in \cite{PipolodeGioia:2022exe, de2023celestial} that AdS-Witten diagrams in $(d+1)$-dimensional spacetime formally reduce to CCFT$_{d-1}$ amplitudes in the limit when operators in the dual Lorentzian CFT on the cylinder are inserted around two antipodally-identified time slices separated by $\Delta \tau =\pi$ in global time. 
Concretely, celestial correlators were proposed to emerge from holographic conformal correlators in one higher dimension according to 
\be 
\label{eq:exp}
\begin{split}
\langle \mathscr{O}^{\eta_1}_{\Delta_1, \lambda_1}(\Omega_1) \cdots \mathscr{O}^{\eta_n}_{\Delta_n, \lambda_n}(\Omega_n) \rangle =  \prod_{i= 1}^n &\left( \Ni^{-1} \int_{-\infty}^{\infty} du_i (u_i + i\eta_i \epsilon)^{-\lambda_i}\right) \\
&\times \left.\lim_{R \rightarrow \infty} \langle {\cal O}_{\Delta_1}(t_1,\Omega_1) \cdots {\cal O}_{\Delta_n}(t_n,\Omega_n)\rangle\right|_{t_i = \eta_i \frac{\pi}{2} + \frac{u_i}{R}}.
\end{split}
\ee 
The integrals over the infinitesimal strips parameterized by $u$ effectively implement a dimensional reduction mapping each CFT$_{d}$ operator ${\cal O}_\Delta(t,\Omega)$ to a continuum of ``KK modes'' $\mathscr{O}^{\eta}_{\Delta, \lambda}(\Omega)$ transforming as primary operators in CCFT$_{d-1}$  upon identifying $\Delta + \lambda - 1$ with the CCFT scaling dimension.
Furthermore, 
\be
\label{eq:normz}
\Ni = \frac{i^{-\Delta} (\eta i)^{-\lambda} \pi^{1 - d/2} R^{\Delta - \frac{d -1}{2}}}{\Gamma(\Delta - \frac{d}{2} + 1) \Gamma(\lambda)}, \quad \eta = \pm 1
\ee
is a normalization factor relating celestial and conformal operators \cite{de2023celestial}. This analysis was extended to spinning operators in \cite{de2023celestial}, where it was then shown that the leading conformally soft gluon theorem, as well as the leading and subleading conformally soft graviton theorems may be recovered from respectively current and stress tensor Ward identities in the higher dimensional CFT. 

In the case of CFT$_3$, an extended BMS$_4$ symmetry was shown to emerge from the conformal symmetry of infinitesimal time strips \cite{de2023celestial}. Remarkably, the emergence of these ``soft'' symmetries was shown to hold for arbitrary CFTs, whether holographic or not, suggesting a plausible connection between CCFT and dimensionally reduced CFT on the Lorentzian cylinder. 
The emergence of extended BMS$_4$ can be seen as a generalization of the standard Inonu-Wigner contraction of the conformal algebra $\mathfrak{so}(2,3)$ to the Poincaré algebra \cite{Inonu}. This is related to, yet appears to be more general than the well-known fact that the conformal Carroll algebra coincides with the BMS algebra \cite{Duval:2014uva}. As a result, the left-hand side of \eqref{eq:exp} is expected to satisfy the constraints of Poincaré symmetry on celestial amplitudes discussed in \cite{Stieberger:2018onx,Law:2019glh}. For simplicity, in the present paper we are going to focus on scalar operators.

There are caveats in rigorously establishing the relation in \eqref{eq:exp}. One assumption which one may worry about is whether the large-$R$ limit commutes with the AdS integrals which perturbatively define the holographic correlators on the RHS. Another reason for concern are contributions to these integrals coming from $O(R)$ regions of AdS space which may lead to divergences rendering the LHS ill defined \cite{komatsu2020landau}. The proposed relation \eqref{eq:exp} between CFT and CCFT correlators also raises more naive conceptual puzzles. For small $n$, celestial amplitudes are known to be  distributional and it is not a priori obvious how these distributions arise from the RHS of \eqref{eq:exp} which involves conventional CFT correlation functions (albeit in a bulk point configuration). Furthermore, the LHS of \eqref{eq:exp} is directly related to on-shell scattering amplitudes and thereby should vanish when all operators are incoming or outgoing (ie. $\eta_i = \pm 1$ are all equal). On the other hand, prior to dimensional reduction, the RHS is non-vanishing even when all operators are placed around the same time slice. 

The goal of this paper is to address some of these questions by establishing \eqref{eq:exp} for two- and three-point functions. Since in these cases the correlators are fixed up to a constant by symmetries, our derivation is general and makes no reference to the bulk. As expected, our results are compatible with the Poincaré symmetry constraints of \cite{Law:2019glh}.

Notably, our analysis demonstrates that the expansion of the two-point function around the two time-slices yields two contributions. One is proportional to a delta function on the $(d-1)$-dimensional sphere, while the other is proportional to the standard CFT$_{d-1}$ two-point structure. Under the assumption that $\Delta \geq \frac{d-1}{2}$, the delta function terms dominate, reproducing the expected behavior of celestial amplitudes. Furthermore, in the case of the two-point function, the integral transforms over the strips accurately reproduce the normalization predicted by the AdS Witten diagram analysis. Keeping careful track of the $i\epsilon$ prescription appearing in the integral transforms, as well as in the definition of the time ordered CFT correlators, we find that the integral transforms in \eqref{eq:exp} project out the CFT correlators involving operators inserted around the same time slices on the cylinder. 

An alternative approach to dimensionally reduce the CFT correlators is to arrange the operators in a bulk-point configuration on the Lorentzian cylinder, shift $\Delta_i \rightarrow \Delta_i^{\rm CCFT} = \Delta_i + \lambda_i - 1$ and take the $u_i\to 0$ limit. From a Witten diagram analysis we expect this procedure to be equivalent to taking the time-Mellin integrals, provided that the integration over AdS spacetime commutes with the integrals over the positions of the operators on the boundary. We demonstrate this explicitly in the case of the two-point function. Remarkably, the same prescription also reproduces the correct Euclidean CCFT three-point function associated to a flat space $\phi^3$ bulk vertex provided one uses the OPE coefficient corresponding to a $\phi^3$ bulk vertex in AdS. 
Obtaining a match requires careful consideration of the distributional components of the Lorentzian three-point function. Renormalizing the operators according to \eqref{eq:normz} is essential to project out the components in the large-$R$ expansion that do not have the correct CCFT three-point structure. In particular, we show in Table \ref{tab:R-scaling} that all other components are suppressed for generic operator dimensions in a \textit{unitary} Lorentzian CFT.
On the other hand, we find that explicit evaluation of the time-Mellin transforms in \eqref{eq:exp}, one recovers the correct result only up to a trigonometric function that can be eliminated by taking the residue at $\sum_i\Delta_i = d+2$. Interestingly, the correct normalization may be directly obtained by reducing the integrals over $u_i$ in \eqref{eq:exp} to an integral over $v \equiv \sum_{i} \omega_i u_i, ~ \omega_i  > 0$ and evaluating this along a certain path inside the universal cover of the complex $v$ plane -- see Appendix \ref{app:3pt-time-mellin}.

The organization of this paper is as follows: Section \ref{sec:preliminaries} establishes conventions and reviews the expected relation between Lorentzian CFT$_d$ correlators and massless celestial amplitudes. Section \ref{sec:celestial-amplitudes} reviews the derivation of Euclidean CCFT$_2$ two- and three-point functions, and provides a generalization of these formulas to Euclidean CCFT$_{d-1}$. Section \ref{sec:two-point} discusses the expansion and dimensional reduction of Lorentzian CFT$_d$ scalar two-point functions, resulting in perfect agreement with CCFT$_{d-1}$ two-point amplitudes of massless scalars. In Section \ref{sec:three-point} we show how to extract CCFT$_{d-1}$  three-point functions from Lorentzian CFT$_d$ correlators. Section \ref{sec:discussion} concludes with a discussion. Several technical details of the computations are provided in the appendices. In Appendix \ref{app:A} we prove two useful delta function identities. In Appendix \ref{app:dim-red} we provide a detailed evaluation of the time-Mellin integrals  in \eqref{eq:exp} for two-point functions and show that they reproduce celestial amplitudes with the correct normalization. In Appendix \ref{sec:three-point-naive} we analyze the leading term in a naive expansion of the Lorentzian three-point correlator around the bulk point limit which is shown to give the structure of a celestial three-point amplitude only for certain integer dimensions of operators in CFT$_d$ that violate unitarity bounds. In Appendix \ref{app:3pt-time-mellin} we provide details of the time-Mellin integrals \eqref{eq:exp} for three-point functions and discuss different choices of contour involved in their evaluation which yield different normalizations of the celestial amplitude.

\section{Preliminaries}\label{sec:preliminaries}

The two-point function of a scalar operator of dimension $\Delta$ in Euclidean CFT$_{d}$ takes the form
\begin{equation}
 \label{eq:two-point}
 \langle {\cal O}_{\Delta}(\tau_1,\Omega_1){\cal O}_{\Delta }(\tau_2,\Omega_2)\rangle_E = \dfrac{2^\Delta C_{\Delta}^{d}}{(-P_1\cdot P_2)^{\Delta}},
\end{equation}
where $P_i$ are embedding space vectors $P_i \in \mathbb{R}^{1,d+1}$ \cite{Penedones:2016voo} and $C_{\Delta}$ is a normalization constant
\begin{equation}
\label{eq:norm}
    C_\Delta^{d} = \dfrac{\Gamma(\Delta)}{2\pi^{d/2}\Gamma(\Delta-\frac{d}{2}+1)}.
\end{equation}
The correlator \eqref{eq:two-point} is a single valued, symmetric function of the positions $(\tau_i,\Omega_i).$

Upon analytic continuation to Lorentzian signature by setting $\tau_i =it_i$, \eqref{eq:two-point} develops a branch cut and a prescription for crossing it needs to be specifed. Upon introducing the $i\epsilon$ prescription 
$\tau_i =i(t_i - i\epsilon_i)$, translation invariance implies that the analytically continued two-point function \eqref{eq:two-point} will be a function of $t_1 - t_2 - i\zeta \epsilon$, where we defined $\zeta = {\rm sgn}(\epsilon_1 - \epsilon_2)$ and $\epsilon = |\epsilon_1 - \epsilon_2|$. Wightman functions are defined to be analytically continued Euclidean correlators with fixed operator orderings, namely \cite{hartman2016causality}
\be 
\label{eq:two-point-Wightman}
\begin{split}
  {\cal W}_{\zeta}(P_1,P_2) = \left. \langle {\cal O}_1(\tau_1,\Omega_1) {\cal O}_2(\tau_2,\Omega_n)\rangle_E \right|_{\tau_i =i(t_i - i\epsilon_i)}.
 \end{split}
\ee
The time orderings $\epsilon_1 > \epsilon_2 > 0$ and $\epsilon_2 > \epsilon_1 >  0$ then correspond to respectively $\zeta  = 1$ and $\zeta = -1.$ To illustrate this, the analytic continuations of the one-dimensional Euclidean correlators
\be 
\langle O_1 O_2 \rangle_{E} = \frac{1}{|\tau_1 - \tau_2|^{2\Delta}},
\ee
are the two Wightman functions
\be 
\begin{split}
 \mathcal{W}_1(P_1, P_2) &= e^{-i\pi \Delta}\frac{1}{(t_{12} - i\epsilon)^{2\Delta}},  \\
  \mathcal{W}_1(P_2, P_1) &= e^{i\pi \Delta}\frac{1}{(t_{12} + i \epsilon)^{2\Delta}} = e^{2\pi i \Delta}\mathcal{W}_{-1}(P_1,P_2)\,, \quad \epsilon >0.
\end{split}
\ee
Note that the Wightman 2-point functions with different time orderings differ by their $i\epsilon$ prescription and by a relative $e^{2\pi i \Delta}$ phase \cite{hartman2016causality}.

We will be interested in correlation functions on the Lorentzian cylinder obtained from \eqref{eq:two-point} by parametrizing 
\be 
\label{eq:LPi}
 P_i = \left(\sinh(\tau_i),\Omega_i,\cosh(\tau_i)\right),
\ee
with $\Omega_i$ the unit normal at a point on $S^{d-1}$, and analytically continuing $\tau_i$ as described above. Using the correspondence between AdS-Witten diagrams and conformal correlators \cite{witten1998anti, freedman1999correlation, penedones2007high}, it was argued in \cite{PipolodeGioia:2022exe, de2023celestial} that CFT$_d$ correlation functions reduce to CCFT$_{d-1}$ amplitudes for particular kinematic configurations of the boundary operators.  Specifically, inserting all operators on infinitesimal time strips $t = \pm \frac{\pi}{2} + \frac{u}{R}$ separated by $\pi$ in the global Lorentzian time $t$, the leading contribution to the AdS bulk-to-boundary propagators for \textit{fixed} operator dimensions was shown to be proportional to a conformal primary wavefunction $\varphi_{\Delta}^{\pm}$
\be 
\label{bulk-to-boundary}
\left. K_{\Delta}(P; X) \right|_{\tau_p = \pm \frac{\pi}{2} + \frac{u}{R}} = \frac{C^d_{\Delta} R^{\Delta - \frac{d-1}{2}}}{i^{\Delta}\Gamma(\Delta)} \varphi^{\pm}_{\Delta}(x - x_0; \hat{q}),
\ee
where $x$ and $\hat{q}$ respectively label points in $\mathbb{R}^{1,d}$ and the celestial $(d-1)$-dimensional space \cite{PipolodeGioia:2022exe},  $x_0 = (u,\vec{0})$, $C_{\Delta}^d$ was defined in \eqref{eq:norm} and 
\be 
\label{eq:cpw}
\varphi_{\Delta}^{\eta}(x;\hat{q}) = \frac{(i \eta )^{\Delta} \Gamma(\Delta)}{(-\hat{q}\cdot x + i\eta\epsilon)^{\Delta}}.
\ee

It was further shown in \cite{de2023celestial} that the dependence on the time coordinate parametrizing the strip can be traded for a shift in the dimension $\Delta$ by the same integral transform relating Carrollian and celestial operators \cite{Donnay:2022aba, Donnay:2022wvx}. Specifically, defining 
\be 
I_{\varphi}^{\pm} \equiv  \int_{-\infty}^{\infty} du (u \pm i\epsilon)^{-\lambda} \varphi^{\pm}_{\Delta}(x - x_0;\hat{q}) = (\pm i)^{\Delta} \Gamma(\Delta) \int_{-\infty}^{\infty} du \frac{(u \pm i\epsilon)^{-\lambda} }{(t - u - r\Omega \cdot \Omega_p \pm i\epsilon)^{\Delta}}
\ee
and evaluating the integral over $u$ keeping careful track of the $i\epsilon$ prescription, we find
\be 
\label{Ipm}
\begin{split}
I_{\varphi}^{\pm} &= (\pm i)^{\Delta} \frac{\pi \Gamma(\Delta+\lambda - 1)}{\sin \pi \Delta \Gamma(\lambda)} \frac{1}{(-\hat{q}\cdot x \pm i \epsilon)^{\Delta + \lambda} |\hat{q}\cdot x|^{-1}}\\
&\times \left[ e^{\mp i \pi \Delta \sgn(-\hat{q}\cdot x)} - \frac{\sin \pi (\Delta + \lambda)}{\sin \pi \lambda} + e^{\mp i\pi \lambda \sgn(-\hat{q}\cdot x)} \frac{\sin \pi \Delta}{\sin \pi \lambda} \right]\\
&= 
2(\pm i)^{\Delta-1} \frac{\pi \Gamma(\Delta + \lambda - 1)}{\Gamma(\lambda)} \frac{1}{(-\hat{q}\cdot x \pm i \epsilon)^{\Delta + \lambda - 1}}. 
\end{split}
\ee
Putting everything together, we have \cite{de2023celestial}
    \begin{equation}
    \label{eq:bulk-bdry-prop-limit}
    \begin{split}
        \lim_{R\to\infty}\int_{-\infty}^\infty du (u\pm i\epsilon)^{-\lambda}K_{\Delta}(P;X)|_{\tau_p = \pm\frac{\pi}{2}+\frac{u}{R}} &= \Ni \varphi^\pm_{\Delta+\lambda-1}(x;\hat{q}),
    \end{split}
\end{equation}
where $\Ni$ is given in \eqref{eq:normz}. From equations \eqref{bulk-to-boundary} and \eqref{Ipm} one can easily infer the relation \eqref{eq:exp} between CCFT$_{d-1}$ amplitudes and CFT$_d$ correlators,  with the normalization factors fixed according to \eqref{eq:normz}. 

Provided that the $R \rightarrow \infty$ limit commutes with the $u$ integrals, \eqref{eq:bulk-bdry-prop-limit} implies that $(d-1)$-dimensional celestial correlators can alternatively be obtained from $d$-dimensional Lorentzian conformal correlators with $\Delta_i\to \Delta_i^{\rm CCFT}=\Delta_i+\lambda_i-1$ and $u_i\to 0$. This coincides with the prescription first proposed in \cite{PipolodeGioia:2022exe} applied to  conformal correlators with shifted dimensions. The analytic continuation in the CFT dimensions should be regarded as a tool to more easily evaluate the time-Mellin integrals. In summary, we expect
\begin{equation}\label{eq:new-prescription}
    \begin{split}
        \lim_{R\to \infty} \prod_{i=1}^n \dfrac{1}{N_{\Delta_i}}\int_{-\infty}^\infty du_i (u_i+i\eta_i\epsilon)^{-\lambda_i}\langle {\cal O}_{\Delta_1}\cdots {\cal O}_{\Delta_n}\rangle &= \lim_{u_i\to 0}\lim_{R\to\infty}\prod_{i=1}^n \frac{i^{\Delta_i^{\rm CCFT}}\Gamma(\Delta_i^{\rm CCFT})}{C^d_{\Delta_i^{\rm CCFT}}R^{\Delta_i^{\rm CCFT} - \frac{d-1}{2}}}\\
    &\times {W}_{\Delta_1^{\rm CCFT}\cdots\Delta_n^{\rm CCFT}}\left(\eta_i\frac{\pi}{2}+\frac{u_i}{R},\eta_i\Omega_i\right),
    \end{split}
\end{equation}
where we denote by $W_{\Delta_1,\dots, \Delta_n}(P_1,\dots, P_n)$ the sum of all $n$-point Witten diagrams with external operators of dimensions $\Delta_i$.

Either way, the derivation of \eqref{eq:exp} involved the representation of CFT$_{d}$ correlators as linear combinations of Witten diagrams in AdS$_{d+1}$. Furthermore, one had to rely on the kinematic limit under consideration commuting with the integrals over AdS. The goal of this paper is to verify \eqref{eq:exp} directly at the level of boundary correlators, with no reference to the bulk. On the LHS of \eqref{eq:exp} we expect to obtain correlators in a \textit{Euclidean} celestial CFT$_{d-1}.$ The calculation of such two and three point correlators is straightforward, although to the extent of our knowledge, explicit formulas in CCFT$_{d-1}$ have not appeared in the literature before with the exception of \cite{Law:2019glh}.\footnote{One instead usually considers the analytic continuation to Lorentzian signature, or the ``celestial torus'', see for example \cite{Pasterski:2017ylz}.} In the next section we review the derivation of Euclidean celestial amplitudes which will be important for explicit comparison with the results obtained from CFT$_d$ later.

\section{Euclidean CCFT$_{d-1}$ amplitudes}

\label{sec:celestial-amplitudes}

Consider first the three-point amplitude for massless scalars of arbitrary helicity in $4$-dimensional Minkowski space. We work in $(3,1)$ bulk signature, which means that we are considering an Euclidean celestial space. The momentum-space three-point amplitudes are completely fixed by the little group symmetry (see eg. \cite{Arkani-Hamed:2017jhn}) and  take the form
\be 
\label{eq:flat-three-point}
\mathcal{M}_{h_1 h_2 h_3} = M(1,2,3) (2\pi)^{4}\delta^{(4)}\left(\sum_{i = 1}^3 \eta_i q_i \right),
\ee
where 
\be 
M(1,2,3) = \begin{cases}
    g [12]^{h_1 + h_2 - h_3} [23]^{h_2 + h_3 - h_1} [31]^{h_3 + h_1 - h_2}, \quad \sum_{i = 1}^3 h_i >0, \vspace{10pt}\\
      g\langle 12\rangle^{-h_1 - h_2 + h_3}  \langle 23\rangle^{-h_2 - h_3 + h_1}  \langle 31 \rangle^{-h_3 - h_1 + h_2}, \quad \sum_{i = 1}^3 h_i < 0.
\end{cases}
\ee 
Here 
\be 
\langle i j\rangle = \lambda_{\alpha} \lambda^{\alpha},\quad [ij] = \tilde{\lambda}^{\dot{\alpha}} \tilde{\lambda}_{\dot{\alpha}}\,,
\ee
and massless momenta $\eta_i q^{\mu}_i$ may be expressed in terms of the spinor helicity variables $\lambda, \tilde{\lambda}$ as
\be 
\eta_i q_{i\mu} \sigma^{\mu}_{\alpha \dot{\alpha}}  \equiv \lambda_{\alpha} \tilde{\lambda}_{\dot{\alpha}}\;,
\ee
where $\sigma^{\mu}$ are Pauli matrices. 
Parametrizing 
\be 
\lambda_i = \omega_i \left(\begin{matrix}
z_i\\
1
\end{matrix}
 \right), \quad \tilde{\lambda}_i = \eta_i\omega_i \left(\begin{matrix}
\bz_i\\
1
\end{matrix}
\right)
\ee
we have
\be 
\langle i j \rangle = \sqrt{\omega_i \omega_j} z_{ij}, \quad [ij] = -\eta_i \eta_j \sqrt{\omega_i \omega_j} \bz_{ij}\;.
\ee
 For a particular choice of basis for the Pauli matrices $\sigma^{\mu}$ (see eg. appendix A of \cite{Arkani-Hamed:2020gyp})
\be 
\label{eq:mom}
q(z,\bz) = \omega \left(1+z\bz, z+\bz, -i(z-\bz), 1 - z\bz \right).
\ee

Using the parametrization \eqref{eq:mom}, we find that the momentum conserving delta function becomes 
\be 
\label{eq:mom-delta-int}
\delta^{(4)}\left(\sum_{i = 1}^{3} \eta_i q_i^{\mu}\right) = \frac{1}{4}\delta\left(\sum_{i} \eta_i \omega_i \right) \delta\left(\sum_i \eta_i \omega_i z_i \bz_i \right) \delta\left(\sum_i \eta_i \omega_i z_i \right)\delta\left(\sum_i \eta_i \omega_i \bz_i \right).
\ee
Solving the last two constraints, we find
\be 
z_1 = -\frac{\eta_2 \omega_2 z_2 + \eta_3 \omega_3 z_3}{\eta_1 \omega_1}
\ee
and similarly for the complex conjugate. Then \eqref{eq:mom-delta-int} becomes
\be 
\label{eq:3-pt-delta}
\begin{split}
\delta^{(4)}\left(\sum_{i = 1}^{3} \eta_i q_i^{\mu}\right) &= \frac{1}{{4}}\delta\left(\sum_{i} \eta_i \omega_i \right) \delta\left(\frac{\eta_2\eta_3 \omega_2 \omega_3 |z_{23}|^2}{\eta_1 \omega_1}\right) \delta\left(\sum_i \eta_i \omega_i z_i \right)\delta\left(\sum_i \eta_i \omega_i \bz_i \right)\\
&= \frac{1}{{4}}\delta\left(\sum_{i} \eta_i \omega_i \right) \delta\left(\frac{\eta_2\eta_3 \omega_2 \omega_3 |z_{23}|^2}{\eta_1 \omega_1}\right) \delta\left( \eta_1 \omega_1 z_{13}\right)\delta\left( \eta_1 \omega_1 \bz_{13} \right)\\
&= \frac{1}{{4}\omega_1 \omega_2 \omega_3} \delta\left(\sum_{i} \eta_i \omega_i \right) \delta\left( |z_{23}|^2 \right) \delta^{(2)}\left( z_{13}\right),
\end{split}
\ee
where in the second line we used the constraint $z_2 = z_3$ imposed by the second delta function in the first line. Following \cite{Law:2019glh} one can further write\footnote{In contrast to \cite{Law:2019glh}, we use the convention $\int_0^{\infty} dr \delta(r) f(r) = f(0)$. } 
\begin{equation}\label{eq:delta-d3}
\delta\left(|z_{23}|^2\right) = \delta(x_{23}^2+y_{23}^2) = \pi    \delta(x_{23})\delta(y_{23}) =\pi \delta^{(2)}(\vec{x}_{23}) = 2\pi \delta^{(2)}(z_{23}).
\end{equation}

Further, assuming $\sum_{i = 1}^3 h_i < 0$ and taking Mellin transforms with respect to the energies $\omega_i$, we find the following 3-point celestial amplitude 
\be 
\label{eq:3-pt-4d}
\begin{split}
\widetilde{A}_3 & \equiv \prod_{i = 1}^3\left(\int_0^{\infty} d\omega_i \omega_i^{\Delta^i_{\rm CCFT} - 1} \right) \mathcal{M}_{h_1 h_2 h_3} \\
&=  {g} |z_{12}|^{-h_1 - h_2 - h_3} e^{-i\delta(h_1 + h_2 + h_3)} \prod_{i = 1}^3 \left(\int_0^{\infty} d\omega_i \omega_i^{\Delta_{\rm CCFT}^i - 1} \right) \omega_1^{-1-h_1} \omega_2^{-1-h_2} \omega_3^{-1-h_3}\delta\left(\sum_{i = 1}^3 \eta_i \omega_i\right) \\
&\times (2\pi)^4 \frac{\pi}{2} \delta^{(2)}({z_{23}})\delta^{(2)}(z_{13}).
 \end{split}
\ee
We can eliminate the delta function in energies by solving for the $\omega_i$ corresponding to an incoming particle ($\eta_i = -1$). Let's choose wlog $\omega_3$ to be incoming and $\omega_1$, $\omega_2$ outgoing. Then
\be
\label{eq:spin}
\begin{split}
\widetilde{A}_3 
&=  {4}^{-1}(2\pi)^6 g |z_{12}|^{-h_1 - h_2 - h_3}  e^{-i\delta(h_1 + h_2 + h_3)} B(\Delta_{\rm CCFT}^2 - h_2 -1, \Delta_{\rm CCFT}^1 - h_1 -1)\\
&\times \delta(\Delta_{\rm CCFT}^1 -h_1+ \Delta_{\rm CCFT}^3 -h_3+ \Delta_{\rm CCFT}^2 -h_2 -4)\delta^{(2)}({z_{23}})\delta^{(2)}(z_{13}).\\
\end{split}
\ee
The other in-out configurations are obtained by permutations. 
Setting the helicities to zero, for scalars we obtain the final results
\begin{equation}
\label{eq:ccft-res}
    \begin{split}
        \widetilde{A}^{\pm 1,\mp 1,\mp 1}_3 &= \frac{1}{ {4}}(2\pi)^6 g B(\Delta_{\rm CCFT}^2-1,\Delta_{\rm CCFT}^3-1)\delta\left(\sum_{i=1}^3\Delta_{\rm CCFT}^i-4\right)\delta^{(2)}({z_{12}})\delta^{(2)}(z_{13}),\\
        \widetilde{A}^{\mp 1,\pm 1, \mp 1}_3 &= \frac{1}{ {4}}(2\pi)^6 g B(\Delta_{\rm CCFT}^1-1,\Delta_{\rm CCFT}^3-1)\delta\left(\sum_{i=1}^3\Delta_{\rm CCFT}^i-4\right)\delta^{(2)}({z_{23}})\delta^{(2)}({z_{12}}),\\
        \widetilde{A}^{\mp 1,\mp 1, \pm 1}_3 &= \frac{1}{ {4}}(2\pi)^6 g B(\Delta_{\rm CCFT}^1-1,\Delta_{\rm CCFT}^2-1)\delta\left(\sum_{i=1}^3\Delta_{\rm CCFT}^i-4\right)\delta^{(2)}({z_{23}})\delta^{(2)}(z_{13}).
    \end{split}
\end{equation}

For completeness, we also give the generalization to higher dimensions of the scalar case which  is straightforward, presenting also the two-point function result for later reference. In the discussion so far we have employed the flat parametrization of the null momenta \eqref{eq:mom} following \cite{Law:2019glh}. To study the dimensional reduction of a CFT$_d$ on the cylinder the round parametrization is more natural and therefore we switch to it now. Note that the three-point coefficients of the celestial amplitudes in both parametrizations differ by factors of $2$.\footnote{More precisely, relating round and planar parametrizations by $q_{\rm round} = \frac{1}{1+|\vec{z}|^2}q_{\rm flat}$ it follows by a change of variables in the energy Mellin transforms that the associated celestial amplitudes of $n$ particles are related by $$\widetilde{A}^{\rm round}_n = \prod_{i=1}^n (1+|\vec{z}_i|^2)^{\Delta_i}\widetilde{A}_n^{\rm flat}.$$
In the case $n=3$, one obtains the following relation between three-point coefficients in the two parametrizations $${\cal C}^{\eta_1\eta_2\eta_3}_{\rm round} = 2^{d+1}{\cal C}_{\rm flat}^{\eta_1\eta_2\eta_3}$$}

 We parametrize on-shell massless momenta in flat $d+1$ dimensional spacetimes by 
\be \label{eq:round-momenta}
q = \omega(1, \Omega), \quad \Omega\cdot \Omega = 1,
\ee
with $\Omega$ the unit normal to $S^{d-1}$.  In this case the celestial in/out two-point function can be straightforwardly evaluated to be
\begin{equation}
    \begin{split}\label{eq:celestial-two-point-round}
        \widetilde{A}_2 &\equiv \int_0^\infty d\omega_1 \omega_1^{\Delta^1_{\rm CCFT}-1}\int_0^\infty d\omega_2\omega_2^{\Delta^2_{\rm CCFT}-1} (2\pi)^{d}2\omega_1\delta^{(d)}(\omega_1 \Omega_1-\omega_2\Omega_2)\\
        &= 2(2\pi)^{d+1} \delta\left(\Delta^1_{\rm CCFT}+\Delta^2_{\rm CCFT}-d+1\right)\delta_{S^{d-1}}(\Omega_1,\Omega_2).
    \end{split} 
\end{equation}

The three-point function associated with the scalar amplitude
\be 
\mathcal{M}_{000} = g  (2\pi)^{d+1} \delta^{(d+1)}\left(\sum_{i = 1}^3 \eta_i q_i \right)
\ee
on the other hand, can be computed by generalizing the calculation we presented in $d=3$ and adapting to the parametrization \eqref{eq:round-momenta}. Using the constraint
\be 
\Omega_1 = \frac{\eta_2\omega_2 \Omega_2 + \eta_3 \omega_3 \Omega_3}{\eta_1 \omega_1} \implies 1 = \frac{\omega_2^{2} + \omega_3^2 + 2\eta_2\eta_3 \omega_2\omega_3 \Omega_2\cdot \Omega_3}{\omega_1^2} \iff \Omega_2\cdot \Omega_3 = 1,
\ee
we find the momentum conserving delta function
\be 
\begin{split}
\delta^{(d+1)}\left(\sum_{i = 1}^3 \eta_i q_i^{\mu} \right) &= {2} \delta\left(\sum_{i} \eta_i \omega_i \right) \delta\left(\frac{2\eta_2\eta_3\omega_2\omega_3}{\eta_1\omega_1}\left(\Omega_2\cdot \Omega_3 - 1 \right)\right) \delta^{(d-1)}\left(\sum_i \eta_i \omega_i \Omega_i\right)\\
&= \frac{1}{\omega_2\omega_3 \omega_1^{d-2}} \delta\left(\sum_{i} \eta_i \omega_i \right)\delta^{(d-1)}\left(\Omega_{13} \right) \delta(\Omega_2\cdot \Omega_3 - 1).
\end{split}
\ee

Now we notice that 
\be 
\begin{split}
\int_{S^{d-1}} f(\Omega_2)\delta(\Omega_2\cdot \Omega_3 - 1) &= \int_{S^{d-2}} \int d\cos \theta_{23} (\sin\theta_{23})^{d-3} f(\Omega_2) \delta(\cos\theta_{23} - 1)  \\
&=\spherevolume{d-2} \lim_{2 \rightarrow 3}(\sin\theta_{23})^{d-3} f(\Omega_3)\;,
\end{split}
\ee
where 
\be 
\spherevolume{d-2} = \frac{2\pi^{\frac{d-1}{2}}}{\Gamma(\frac{d-1}{2})}\;.
\ee
Therefore
\be 
\label{eq:delta-fct-mom}
\begin{split}
\delta(\Omega_2\cdot \Omega_3 - 1) &= \spherevolume{d-2} (\sin\theta_{23})^{d-3} \delta^{(d-1)}(\Omega_3 - \Omega_2)\\
&= \spherevolume{d-2}  |\Omega_2 - \Omega_3|^{d-3} \delta^{(d-1)}(\Omega_2 - \Omega_3).
\end{split}
\ee

Then we immediately find the generalization of \eqref{eq:ccft-res} to higher dimensions 
\be 
\begin{split}\label{eq:ccft-res-d}
 \widetilde{A}^{\pm 1,\mp 1,\mp 1}_{3} &= {2}{(2\pi)^{d+1}}\frac{2\pi^{\frac{d+1}{2}}g}{\Gamma(\frac{d-1}{2})} B(\Delta_{\rm CCFT}^2-1,\Delta_{\rm CCFT}^3- 1){|\Omega_{23}|^{d-3}}\delta^{(d-1)}({\Omega_{12}})\delta^{(d-1)}(\Omega_{13})\delta_{\Delta} ,\\
        \widetilde{A}^{\mp 1,\pm 1, \mp 1}_{3} &= {2}{(2\pi)^{d+1}}\frac{2\pi^{\frac{d+1}{2}}g}{\Gamma(\frac{d-1}{2})} B(\Delta_{\rm CCFT}^1-d+2,\Delta_{\rm CCFT}^3-1){|\Omega_{23}|^{d-3}}\delta^{(d-1)}(\Omega_{23})\delta^{(d-1)}({\Omega_{12}})\delta_{\Delta} ,\\
        \widetilde{A}^{\mp 1,\mp 1, \pm 1}_{3} &={2}{(2\pi)^{d+1}}\frac{2\pi^{\frac{d+1}{2}}g}{\Gamma(\frac{d-1}{2})}  B(\Delta_{\rm CCFT}^1-d+2,\Delta_{\rm CCFT}^2-1) {|\Omega_{23}|^{d-3}}\delta^{(d-1)}(\Omega_{23})\delta^{(d-1)}(\Omega_{13})\delta_{\Delta} ,
\end{split}
\ee
where we defined
\be 
\delta_{\Delta} \equiv \delta\left(\sum_{i=1}^3\Delta_{\rm CCFT}^i-d-1\right) \;.
\ee
In the next section we show that \eqref{eq:ccft-res-d} can be obtained by considering the kinematic limits of CFT$_{d}$ correlators described in Section \ref{sec:preliminaries}. 

\section{Two-point function}
\label{sec:two-point}

We start with the Wightman two-point function of scalars in CFT$_d$ on the Lorentzian cylinder
\begin{equation}
\label{eq:wightman-twopt}
\begin{split}
    {\cal W}_\zeta(P_1,P_2) &\equiv \dfrac{2^\Delta C^d_\Delta}{(\cosh\left(\zeta i t_{12} + \epsilon\right)-\Omega_1\cdot \Omega_2)^\Delta},\quad \zeta =\pm 1,
    \end{split}
\end{equation}
where $P_i$ are defined in \eqref{eq:LPi}, $C^d_{\Delta}$ is the normalization in \eqref{eq:norm} and $\zeta$ determines the operator ordering according to \eqref{eq:two-point-Wightman}. We now demonstrate that CCFT$_{d-1}$ two-point functions are obtained from \eqref{eq:wightman-twopt} following the prescription outlined in the previous section. We break this down into the following two steps:
\begin{enumerate}
\item Set $t_i=\eta_i\frac{\pi}{2}+\frac{u_i}{R}$ and $\Omega_i \rightarrow \eta_i \Omega_i$, $\eta_i = \pm 1$ and take $R \rightarrow \infty.$ 
\item Transform the result to a $(d-1)$-dimensional conformal primary basis by integrating each external leg against $\int_{-\infty}^{\infty} du_i (u_i + i\eta_i \epsilon)^{-\lambda_i}$.
\end{enumerate}

The first point is expected on general grounds to yield a $d$-dimensional Carrollian correlator from a AdS$_{d+1}$ Witten diagram \cite{Bagchi:2023fbj}, while the second point implements the dimensional reduction to a celestial $(d-1)$-dimensional correlator \cite{de2023celestial}. The two key outcomes of the prescription which we now demonstrate are the localization of \eqref{eq:wightman-twopt} to a delta function in the transverse $\Omega$ directions to leading order as $R \rightarrow \infty$ and the relation \eqref{eq:exp} for $n = 2$
provided $\Ni$ are chosen according to \eqref{eq:normz}.

\subsection{Distributions on S$^{d-1}$  from kinematic limits of CFT$_{d}$ two-point functions}

We start by discussing the large-$R$ expansion of \eqref{eq:wightman-twopt} for the kinematic configuration in point 1 above. For $u_1 \neq u_2$, the large-$R$ expansion of \eqref{eq:wightman-twopt} contains both polynomial and delta-function singular terms as $z_1 \rightarrow z_2$. We will show in Section \ref{sec:two-point-norm} that the $u_i$ integrals project out the configurations where $\eta_1\eta_2 = 1$ corresponding to operators inserted in the same strip. We therefore focus on the case $\eta_1 \eta_2 = -1.$ Using the identity \cite{Kutasov:1999xu}
\begin{equation}\label{eq:dirac-delta-expansion}
\begin{split}
    \left(\dfrac{\epsilon}{\epsilon^2+z^2}\right)^\Delta &=\left[\dfrac{\pi^{n/2}\Gamma\left(\Delta-\frac{n}{2}\right)}{\Gamma(\Delta)\epsilon^{\Delta-n}}\delta^{(n)}(z)+O(\epsilon^{n-\Delta+2})\right]+\left[\dfrac{\epsilon^\Delta}{z^{2\Delta }}+O(\epsilon^{\Delta+2})\right], \quad z\in\mathbb{R}^n,
    \end{split}
\end{equation}
which we prove in Appendix \ref{app:A}, we find the two leading contributions to \eqref{eq:wightman-twopt} as $R \rightarrow \infty$ 
\begin{equation}\label{eq:2pt-expansion}
    \begin{split}
        \dfrac{1}{(-P_1\cdot P_2 + i{\eta_1 \zeta}\epsilon)^\Delta} &= 2^\Delta e^{-i\pi \Delta{\eta_1\zeta} \theta(-\eta_1\eta_2)}\\
        &\times \bigg[\bigg(\frac{C_1\delta_{S^{d-1}}(z_1,z_2)}{(u_{12} - i\zeta \varepsilon)^{2\Delta - d+1}} +O(R^{2\Delta-d-1})\bigg)+ \bigg(\dfrac{ C_2}{(\Omega_1-\Omega_2)^{2\Delta}}+O(R^{-2})\bigg)\bigg],
    \end{split}
\end{equation}
where 
\be 
\label{eq:c12}
\begin{split}
C_1 = \dfrac{R^{2\Delta-d+1}\pi^{\frac{d-1}{2}}(-i\zeta)^{2\Delta - d+1}\Gamma(\Delta-\frac{d-1}{2})}{\Gamma(\Delta)}, \quad C_2 = 1,
\end{split}
\ee
and $\theta$ is the unit step function. 

Under the condition $\Delta > \frac{d-1}{2}$ which is almost always\footnote{The unitarity bound for scalars in CFT$_{d}$ is $\Delta \geq \frac{d}{2} - 1$ \cite{Simmons-Duffin:2016gjk}. The case $\frac{d}{2} - \frac{1}{2} \geq \Delta \geq \frac{d}{2}-1$ requires a separate analysis (see also \cite{Kutasov:1999xu}) which we postpone to future work. } obeyed in unitary CFT$_d$, the second series of terms in \eqref{eq:2pt-expansion} is subleading. Under this assumption, we find the leading term in the large-$R$ expansion of the Wightman function \eqref{eq:wightman-twopt} with  $t_i=\eta_i\frac{\pi}{2}+\frac{u_i}{R}$
\begin{equation}
\label{eq:two-point-flat}
    \begin{split}
        \dfrac{2^\Delta C^d_{\Delta}}{(-P_1\cdot P_2 {+i\eta_1\zeta\epsilon})^{\Delta}} &= \dfrac{\widetilde{C}^d_\Delta}{|u_1 - u_2|^{2\Delta - d+1}}\delta_{S^{d-1}}(z_1,z_2),
    \end{split}
\end{equation}
where we defined
\begin{equation}
\label{eq:C-tilde}
    \begin{split}
        \widetilde{C}^d_\Delta 
  &= \dfrac{e^{-i\pi \Delta{\eta_1\zeta}\theta(-\eta_1\eta_2)- i\frac{\pi}{2} \zeta(2\Delta - d + 1)\sgn{\left(u_{12} \right)}}\Gamma(2\Delta- d+1)}{2^{1-d} R^{-2\Delta+d-1}\Gamma(\Delta-\frac{d}{2}+1)^2}
    \end{split}
\end{equation}
 and we used the Legendre duplication formula to simplify the result. Since \eqref{eq:two-point-flat} including the phases will be crucial moving forward, we next provide another derivation using a series representation of the CFT$_d$ two-point function.  
 
\subsubsection{Alternative derivation of the expansion \eqref{eq:2pt-expansion}}

In this section we give an alternative calculation of the large $R$ expansion of the CFT$_d$ two-point function by employing a series representation. This version of the calculation is set from the start in Lorentzian signature, thereby evading the need of analytic continuation of Euclidean CFT correlators. The series representation we are going to use can be found by considering QFT in Lorentzian AdS$_{d+1}$. Note, however, that since conformal symmetry fixes two-point functions up to normalization, one ends up obtaining a representation of arbitrary conformal two-point functions in Lorentzian signature.

Let $\phi(x)$ be a free scalar field in AdS with mass $m^2R^2=\Delta(\Delta-d)$ where $\Delta$ is a given parameter, to be identified with the dimension of the an operator in the dual CFT. We are going to exploit the \textit{extrapolate dictionary}, which states that CFT correlators are boundary values of bulk correlators, or more precisely
\begin{equation}
    \langle {\cal O}(t_1,\Omega_1)\cdots{\cal O}(t_n,\Omega_n)\rangle_{\rm CFT} = \lim_{\rho_i \to \frac{\pi}{2}} \prod_{i=1}^n(\cos\rho_i)^{-\Delta}\langle \phi(t_1,\rho_1,\Omega_1)\cdots \phi(t_n,\rho_n,\Omega_n)\rangle_{\rm AdS},
\end{equation}
{where $(t,\rho,\Omega)$ are global coordinates in AdS$_{d+1}$.}

The bulk correlators can be computed in perturbation theory according to the rules of QFT in curved spacetimes after imposing boundary conditions at the timelike boundary. We will assume the standard Dirichlet boundary conditions. As such, a free scalar field in AdS admits the mode expansion
\begin{equation}
    \phi(x)=\sum_{n,J,m_i}\left(e^{-iE_{n,J}t}R_{n,J}(\rho)Y_{J}^{m_i}(\Omega)a_{n,J,m_i}+\text{h.c.}\right),
\end{equation}
where $E_{n,J}$ are the energies \cite{ishibashi2004dynamics,li2021notes}
\begin{equation}
\label{eq:E}
    E_{n,J}=\Delta+J+2n,
\end{equation}
$Y_J^{m_i}(\Omega)$ are $(d-1)$-dimensional spherical harmonics and $R_{n,J}(\rho)$ are the radial wavefunctions
\begin{equation}
    R_{n,J}(\rho) = \dfrac{1}{N_{\Delta,J}}(\sin\rho)^J (\cos\rho)^\Delta {_2F_1}\left(-n,\Delta+J+n;J+\frac{d}{2};\sin^2\rho\right).
\end{equation}
The normalization $N_{\Delta,J}$ is fixed by imposing the canonical commutation relations
\begin{equation}
    [a_{h,J,m_i},a^\dagger_{n',J',m_i'}] = \delta_{n,n'}\delta_{J,J'}\delta_{m_i,m_i'},
\end{equation}
and it reads \cite{li2021notes}\footnote{Our $N_{\Delta,J}$ differs from the one in \cite{li2021notes} by a factor of $2^{2\Delta}$. That happens because our choice of normalization of the two-point function is different, namely $$\langle{\cal O}(t_1,\Omega_1){\cal O}(t_2,\Omega_2)\rangle_{\text{here}} = 2^{2\Delta} \langle{\cal O}(t_1,\Omega_1){\cal O}(t_2,\Omega_2)\rangle_{\text{there}}.$$ See e.g. equations B.3 and 3.11 in \cite{li2021notes}.}
\begin{equation}\label{eq:radial-wf-normalization}
    N_{\Delta,J} =  \sqrt{\dfrac{n!\Gamma(J+\frac{d}{2})^2\Gamma(\Delta+n-\frac{d-2}{2})}{2^{2\Delta}\Gamma(n+J+\frac{d}{2})\Gamma(\Delta+n+J)}}.
\end{equation}

The AdS two-point Wightman function can be obtained immediately as
\begin{equation}
    \begin{split}
        \langle \vac |\phi(x_1)\phi(x_2)|\vac\rangle &= \sum_{n,J,m_i}e^{-i E_{n,J}t_{12}}R_{n,J}(\rho_1)\bar{R}_{n,J}(\rho_2)Y_{J}^{m_i}(\Omega_1)\bar{Y}_J^{m_i}(\Omega_2),
    \end{split}
\end{equation}
where we abbreviate $t_{12}=t_1-t_2$. This sum has been explicitly performed in \cite{burgess1985propagators, inami1985one, giddings2000flat} to yield the bulk-to-bulk propagator in Lorentzian signature. Here, instead, we use it directly to evaluate the boundary correlator using the extrapolate dictionary. The important part of this computation is that the radial wavefunction has the near-boundary behavior
\begin{equation}
    \lim_{\rho\to\frac{\pi}{2}}(\cos\rho)^{-\Delta} R_{n,J}(\rho) = \dfrac{1}{N_{\Delta,J}}\dfrac{\Gamma(\frac{d}{2}-\Delta)\Gamma(\frac{d}{2}+J)}{\Gamma(\frac{d}{2}+J+n)\Gamma(\frac{d}{2}-n-\Delta)}.
\end{equation}
Using \eqref{eq:radial-wf-normalization} we find
\begin{equation}
\begin{split}
   & \lim_{\rho\to\frac{\pi}{2}}(\cos\rho)^{-\Delta} R_{n,J}(\rho) = N_{\Delta,n,J}^{\cal O},\\
    &N_{\Delta,n,J}^{\cal O}\equiv \sqrt{\dfrac{2^{2\Delta}\Gamma(\Delta+n+J)\Gamma(1-\frac{d}{2}+\Delta+n)}{\Gamma(n+1)\Gamma(\frac{d}{2}+J+n)}}\dfrac{{(-1)^n}}{\Gamma(1-\frac{d}{2}+\Delta)},
    \end{split}
\end{equation}
and therefore the two-point function reads
\begin{equation}
\label{eq:two-pt-series}
    \begin{split}
        \langle \vac |{\cal O}(t_1,\Omega_1){\cal O}(t_2,\Omega_2)|\vac \rangle &= \sum_{n,J,m_i}\left(N_{\Delta,n,J}^{\cal O}\right)^2e^{-i E_{n,J}t_{12}}Y_{J}^{m_i}(\Omega_1)\bar{Y}_J^{m_i}(\Omega_2).
    \end{split}
\end{equation}
This gives a series representation of the Wightman  two-point  function, which should hold regardless of existence of a bulk dual. Note, in particular, that the opposite ordering is related to \eqref{eq:two-pt-series} by complex conjugation:\footnote{This can be shown directly using the operator formalism: $$ \langle \vac|\phi(x_1)\phi(x_2)|\vac\rangle^\ast = \langle \vac|\phi^\dagger(x_2)\phi^\dagger(x_1)|\vac\rangle = \langle\vac|\phi(x_2)\phi(x_1)|\vac\rangle,$$ where real scalars have been assumed in the last step.}
\begin{equation}
    \begin{split}
        \langle \vac |{\cal O}(t_2,\Omega_2){\cal O}(t_1,\Omega_1)|\vac \rangle &= \sum_{n,J,m_i}\left(N_{\Delta,n,J}^{\cal O}\right)^2e^{i E_{n,J}t_{12}}\bar{Y}_{J}^{m_i}(\Omega_1){Y}_J^{m_i}(\Omega_2).
    \end{split}
\end{equation}
To dimensionally reduce we now set $t_{12}=\frac{u_{12}}{R}+(\eta_1-\eta_2)\frac{\pi}{2}$. We also replace $\Omega_i\to \eta_i\Omega_i$ to implement the antipodal matching. In this case we obtain the two-point function
\begin{equation}
    \begin{split}
        \langle \vac |{\cal O}(t_1,\Omega_1){\cal O}(t_2,\Omega_2)|\vac \rangle &= \sum_{n,J,m_i}\left(N_{\Delta,n,J}^{\cal O}\right)^2e^{-i \frac{E_{n,J}}{R}u_{12}}e^{-i E_{n,J}(\eta_1-\eta_2)\frac{\pi}{2}}Y_{J}^{m_i}(\eta_1\Omega_1)\bar{Y}_J^{m_i}(\eta_2\Omega_2).
    \end{split}
\end{equation}

Before proceeding we note that the phase $e^{-iE_{n,J}(\eta_1-\eta_2)\frac{\pi}{2}}$ vanishes if $\eta_1=\eta_2$. On the other hand, when $\eta_2=-\eta_1$
\begin{equation}
    e^{-iE_{n,J}(\eta_1-\eta_2)\frac{\pi}{2}R} = e^{-i\eta_1 E_{n,J}\pi} = e^{-i\eta_1\pi \Delta}e^{-i\eta_1\pi J}e^{-i\eta_{1} 2n\pi } = e^{-i\eta_1\pi \Delta}e^{-i\eta_1\pi J}.
\end{equation}
We can then use $e^{-i\eta_1\pi J}$ to flip the sign inside of one of the spherical harmonics
\begin{equation}
    e^{-i\eta_1\pi J} \bar Y^{m_i}_J(\eta_2\Omega_2) =  \bar Y^{m_i}_J(-\eta_2\Omega_2) = \bar Y^{m_i}_J(\eta_1\Omega_2).
\end{equation}
Since $Y_{J}^{m_i}(\eta_1\Omega_1)\bar{Y}_J^{m_i}(\eta_1\Omega_2) = Y_{J}^{m_i}(\Omega_1)\bar{Y}_J^{m_i}(\Omega_2)$ we conclude that
\begin{equation}
\label{eq:int-dec}
    \begin{split}
        \langle \vac |{\cal O}(t_1,\Omega_1){\cal O}(t_2,\Omega_2)|\vac \rangle &= e^{-i\pi\eta_1\Delta \theta(-\eta_1\eta_2)}\sum_{n,J,m_i}\left(N_{\Delta,n,J}^{\cal O}\right)^2e^{-i \frac{E_{n,J}}{R}u_{12}}Y_{J}^{m_i}(\Omega_1)\bar{Y}_J^{m_i}(\Omega_2).
    \end{split}
\end{equation}

We can now decompose the sum \eqref{eq:int-dec} into terms where $E_{n,J}$ scales with $R$ and terms where it does not. The $O(R)$ terms dominate and we consider them first. For these terms we must have $E_{n,J}=\omega R$ which combined with \eqref{eq:E} can be used to promote this sum to an integral. Setting $\omega_n = \frac{2n+\Delta+J}{R}$ we have
\begin{equation}
    \Delta\omega_n \equiv \omega_{n+1} - \omega_n = \frac{2}{R}.
\end{equation}
Eliminating $n$ in the sum in favor of $\omega_n$ we have a sum of the form $\sum_{n}f(\omega_n)$ for some function $f(\omega)$. We can then write
\begin{equation}
    \sum_n f(\omega_n) = \frac{R}{2}\sum_{n} f(\omega_n) \Delta\omega_n
\end{equation}
and take $R\to \infty$, which also takes $\Delta\omega_n\to 0$. We see that in the limit the sum can be interpreted as the Riemann integral of $f(\omega)$ over $[0,+\infty)$. This means that we may identify \cite{li2021notes, Hijano:2019flat}
\begin{equation}
    \sum_{n=0}^\infty \to \dfrac{R}{2}\int_0^\infty d\omega.
\end{equation}
It is also immediate to evaluate the large-$R$ limit of the normalization factors. Using Stirling's approximation of the Gamma functions we obtain
\begin{equation}
    \begin{split}
        \left(N_{\Delta,n,J}^{\cal O}\right)^2 &\simeq  \left(R\omega\right)^{2\Delta-d}\dfrac{2^{d}}{\Gamma(1-\frac{d}{2}+\Delta)^2}.
    \end{split}
\end{equation}

Putting everything together we arrive at the large-$R$ formula
\begin{equation}
    \begin{split}
        \langle \vac |{\cal O}(t_1,\Omega_1){\cal O}(t_2,\Omega_2)|\vac \rangle &\simeq \dfrac{2^{d}e^{-i\pi\eta_1\Delta\theta(-\eta_1\eta_2) }R^{1+2\Delta-d}}{2\Gamma(1-\frac{d}{2}+\Delta)^2}\\
        &\times \int_0^\infty d\omega \omega^{2\Delta-d} e^{-i\omega u_{12}}\sum_{J,m_i}Y_{J}^{m_i}(\Omega_1)\bar{Y}_J^{m_i}(\Omega_2).
    \end{split}
\end{equation}
Evaluating the Mellin integral and using the completeness relation for the spherical harmonics, we finally obtain
\begin{equation}\label{eq:2pt-large-R-same-slice-2}
    \begin{split}
        \langle \vac |{\cal O}(t_1,\Omega_1){\cal O}(t_2,\Omega_2)|\vac\rangle  & \simeq \dfrac{2^{d-1}e^{-i\pi\eta_1\Delta\theta(-\eta_1\eta_2)}R^{1+2\Delta-d}}{\Gamma(1-\frac{d}{2}+\Delta)^2}\dfrac{(-i)^{2\Delta-d+1}\Gamma(2\Delta-d+1)}{(u_{12}-i\epsilon)^{2\Delta-d+1}}\\
      & \times \delta_{S^{d-1}}(\Omega_1,\Omega_2),
    \end{split}
\end{equation}
where we note that the $i\epsilon$ prescription is fixed by convergence of the Mellin transform. The other ordering can be obtained by complex conjugation.

We can summarize both results with a parameter $\zeta=\pm 1$, that is positive for the ${\cal O}(t_1,\Omega_1){\cal O}(t_2,\Omega_2)$ ordering and negative for the other ordering:
\begin{equation}
    {\cal W}_{\zeta} = \dfrac{2^{d-1}e^{-i\pi\zeta\eta_1\Delta \theta(-\eta_1\eta_2)}R^{1+2\Delta-d}}{\Gamma(1-\frac{d}{2}+\Delta)^2}\dfrac{(-\zeta i)^{2\Delta-d+1}\Gamma(2\Delta-d+1)}{(u_{12}-i\zeta \epsilon)^{2\Delta-d+1}}\delta_{S^{d-1}}(\Omega_1,\Omega_2)+\cdots.
\end{equation}
The subleading terms now come from the terms of the series where $E_{n,J}$ does not scale with $R$. For those terms we can approximate $e^{-i \frac{E_{n,J}}{R}u_{12}}\simeq 1$ and we see that we are effectively computing the two-point function at equal times. This coincides with the CFT$_d$ Euclidean two-point function at equal Euclidean times and therefore gives the expansion of Lorentzian CFT$_d$ two-point functions with leading and subleading terms as 
\begin{equation}\label{eq:2pt-expansion-complete}
    \begin{split}
        {\cal W}_{\zeta} &= \dfrac{2^{d-1}e^{-i\pi\zeta\eta_1\Delta \theta(-\eta_1\eta_2)}R^{1+2\Delta-d}}{\Gamma(1-\frac{d}{2}+\Delta)^2}\dfrac{(-\zeta i)^{2\Delta-d+1}\Gamma(2\Delta-d+1)}{(u_{12}-i\zeta \epsilon)^{2\Delta-d+1}}\delta_{S^{d-1}}(\Omega_1,\Omega_2)\\
        &+ e^{-i\zeta\eta_1\pi\Delta \theta(-\eta_1\eta_2)}\dfrac{2^{2\Delta}C^d_\Delta}{|\Omega_1-\Omega_2|^{2\Delta}}+\cdots 
    \end{split}
\end{equation}
in perfect agreement with \eqref{eq:2pt-expansion}.

\subsection{Celestial two-point normalization}
\label{sec:two-point-norm}
In this section we implement the second step of our prescription and demonstrate that the result matches perfectly with the CCFT$_{d-1}$ two-point function. Remarkably, the integral transforms over the strips project out the two-point functions corresponding to operators inserted around the same strip, as we now show. 

Translationally invariant functions $f(u_1, u_2)$ are put in a CCFT$_{d-1}$ conformal primary basis by the integral transform \cite{Pasterski:2021dqe, de2023celestial} 
\begin{equation}
\label{eq:time-Mellin-integral-main}
\begin{split}
    {\cal I}_{\eta_1\eta_2}^{\lambda_1\lambda_2}[f] &\equiv \int_{-\infty}^{\infty} du_1 (u_1+i\eta_1\epsilon)^{-\lambda_1}\int_{-\infty}^\infty du_2 (u_2+i\eta_2\epsilon)^{-\lambda_2} f(u_1-u_2).
   \end{split}
\end{equation}
We show in Appendix \ref{app:dim-red} that under a change of variables, \eqref{eq:time-Mellin-integral-main} becomes
\begin{equation}
\label{eq:int}
\begin{split}
    {\cal I}^{\lambda_1\lambda_2}_{\eta_1\eta_2}[f] &= I^{\sigma}_{\eta_1\eta_2}(\lambda_1,\lambda_2)\int_{-\infty}^\infty dx |x|(x+i\eta_1\epsilon)^{-\lambda_1}(x+i\eta_2\epsilon)^{-\lambda_2}f(x),
    \end{split}
\end{equation}
where 
\begin{equation} \label{eq:2pt-integral-result-main}
    \begin{split}
I_{\eta_1\eta_2}^\sigma(\lambda_1,\lambda_2) &= 2\pi i\dfrac{ \Gamma(\lambda_1+\lambda_2-1)}{\Gamma(\lambda_1)\Gamma(\lambda_2)}\times \begin{cases}
            0,& \eta_1\eta_2 = 1\\
            -\sigma\eta_1 e^{-i\pi\sigma\eta_2\lambda_2},& \eta_1 \eta_2 = -1.
        \end{cases}
    \end{split}
\end{equation}
We see that the dimensionally reduced Wightman two-point functions with operators on the same time strip vanish. This is consistent with the vanishing of celestial amplitudes with operators in in/in or out/out configurations.

Further, performing the integral  \eqref{eq:2pt-integral-result-main} with $f$ in \eqref{eq:two-point-flat} we find  (see Appendix \ref{app:dim-red})
\be \label{eq:2pt-wightman-functions-final}
\begin{split}
 {\cal I}^{\lambda_1\lambda_2}_{\eta_1\eta_2}[{\cal W}_\zeta]&= e^{-i\pi \Delta} e^{\frac{i\pi}{2}\eta_1(\lambda_2 - \lambda_1)}  \frac{8 \pi^3 2^{d-1} R^{2\Delta - d + 1} }{\Gamma(\lambda_1)\Gamma(\lambda_2) \Gamma(\Delta - \frac{d}{2} + 1)^2} \\
 &\times \delta_{S^{d-1}}(z_1,z_2)\delta(\Delta_{\rm CCFT}^1+\Delta_{\rm CCFT}^2 -d+1),\quad \eta_1=\zeta
 \end{split}
\ee
and $ {\cal I}^{\lambda_1\lambda_2}_{\eta_1\eta_2}[{\cal W}_\zeta] = 0$  for $\eta_1 = -\zeta.$ Here we defined the CCFT dimensions \cite{de2023celestial}
\be 
\label{eq:dccft}
\Delta_{\rm CCFT} = \Delta + \lambda-1.
\ee

 We can also consider the projection of the CFT$_d$ time-ordered two-point correlator to the CCFT$_{d-1}$. The CFT$_d$ time-ordered correlator can be expressed as a linear combination of Wightman functions
\begin{equation}
\label{eq:time-ordered-2pt}
    \langle \vac|{\cal T}\{{\cal O}_1(t_1,\Omega_1){\cal O}_2(t_2,\Omega_2)\}|\vac\rangle = {\cal W}_+ \theta(t_{12})+{\cal W}_{-}\theta(t_{21}). 
\end{equation}
After parameterizing $t_i = \eta_i \frac{\pi}{2}+\frac{u_i}{R}$, the step functions admit the following large $R$ limit
\begin{equation}
\label{eq:theta-exp}
    \lim_{R\to \infty}\theta(t_{ij}) = \begin{cases}1,& \eta_i=1,\eta_j=-1,\\
    0,&\eta_i=-1,\eta_j=1\\
\theta(u_{ij}),&\eta_i=\eta_j.
\end{cases}
\end{equation}
This identity can be rigorously proved in the distributional sense. Let $\ell_{ij}=(\eta_i-\eta_j)\frac{\pi}{2}$ and consider $\theta(\frac{ u_{ij}}{R}+\ell_{ij})$ as a distribution acting on functions $f(u_i)$. It is immediate to see that
\begin{equation}
    \lim_{R\to\infty}\int_{-\infty}^\infty du_i f(u_i)\theta\left(\frac{u_{ij}}{R}+\ell_{ij}\right) = \lim_{R\to\infty}\int_{x_j-R\ell_{ij}}^\infty f(u_i)=\begin{cases}1,& \ell_{ij}>0,\\
    0,&\ell_{ij}<0,\\
    \int_{-\infty}^\infty f(u_i)\theta(u_{ij}),& \ell_{ij}=0.\end{cases}
\end{equation}
 Therefore, for $\eta_1=-\eta_2=1$ only the first term in \eqref{eq:time-ordered-2pt} survives, while for $\eta_1=-\eta_2=-1$ only the second term survives. In either case, the coefficients precisely reduce  to the non-vanishing Wightman functions \eqref{eq:2pt-wightman-functions-final}. We conclude that after accounting for the normalization \eqref{eq:normz}, the time-ordered CFT$_d$ two-point function reduces precisely to the scalar two-point function in CCFT$_{d-1}$ given by \eqref{eq:celestial-two-point-round}, namely,
 \be 
 \label{eq:two-point-final}
\langle \mathscr{O}^{\eta_1}_{\Delta, \lambda_1} \mathscr{O}^{\eta_2}_{\Delta, \lambda_2} \rangle = 2 (2\pi)^{d+1}   \delta_{\eta_1, -\eta_2} \delta(\Delta_{\rm CCFT}^1+\Delta_{\rm CCFT}^2 -d+1)\delta_{S^{d-1}}(z_1,z_2).
\ee

One can bypass the subtleties encountered when computing the $u_i$ integrals in \eqref{eq:time-Mellin-integral-main} by directly applying \eqref{eq:new-prescription}. This is done in Appendix \ref{app:dim-reduction-ui->0} which allows us to again precisely recover \eqref{eq:two-point-final}.  

\section{Three-point function}\label{sec:three-point}

Next we consider scalar three-point functions. In Euclidean CFT$_d$ they take the form
\begin{equation}
\label{eq:3-pt}
    \langle {\cal O}_{1}(P_1){\cal O}_{2}(P_2){\cal O}_{3}(P_3)\rangle_E = \frac{C_{123} }{(-P_1\cdot P_2)^{\frac{\alpha_{12}}{2}}(-P_1\cdot P_3)^{\frac{\alpha_{13}}{2}}(-P_2\cdot P_3)^{\frac{\alpha_{23}}{2}}},
\end{equation}
with
\begin{equation}
\begin{split}
    \alpha_{12} &= \Delta_1+\Delta_2-\Delta_3
    \end{split}
\end{equation}
and $\alpha_{13}, \alpha_{23}$ obtained by permutation. Motivated by our analysis of the two-point function, we consider a time-ordered three-point function defined with the following $i\epsilon$ prescription \cite{Kravchuk:2018htv}\footnote{This $i\epsilon$ prescription is not a valid representation of the time-ordered three-point function away from the region $-\pi < \tau_i < \pi$ to which our operators are restricted to. For more general configurations the operators can be related to their images in this region by the operator ${\cal T}$ defined in \cite{Kravchuk:2018htv} which introduces additional phases (see e.g. Eq. 2.15 in that reference).}
\begin{equation}
\label{eq:euclidean-three-point}
    \langle{\cal O}_1(P_1){\cal O}_2(P_2){\cal O}_3(P_3)\rangle = \dfrac{C_{123}}{(-P_1\cdot P_2+i\epsilon)^{\frac{\alpha_{12}}{2}}(-P_1\cdot P_3+i\epsilon)^{\frac{\alpha_{13}}{2}}(-P_2\cdot P_3+i\epsilon)^{\frac{\alpha_{23}}{2}}}.
\end{equation}
In holographic CFTs, a Witten diagram computation shows that the OPE coefficient is expected to scale with $R$ as $C_{123}=R^{\frac{5-d}{2}}\hat{C}_{123}$, where $\hat{C}_{123}$ is independent of $R$.\footnote{Starting with the formula for the bulk-to-boundary propagator \eqref{bulk-to-boundary}, we have \cite{freedman1999correlation}
\begin{equation}\label{eq:ads-3pt-witten-diagram}
    \langle {\cal O}_{\Delta_1}(P_1){\cal O}_{\Delta_2}(P_2){\cal O}_{\Delta_3}(P_3)\rangle = \dfrac{1}{R^{\frac{3(d-1)}{2}}}\int_{\rm AdS} d^{d+1}X \dfrac{1}{(-P_1\cdot \frac{X}{R})^{\Delta_1}}\dfrac{1}{(-P_2\cdot \frac{X}{R})^{\Delta_2}}\dfrac{1}{(-P_3\cdot \frac{X}{R})^{\Delta_3}}.
\end{equation}
Changing variables to $Y= \frac{X}{R}$ we obtain an integral over an AdS space of unit radius, which allows us to establish that
\begin{equation}
    \langle {\cal O}_{\Delta_1}(P_1){\cal O}_{\Delta_2}(P_2){\cal O}_{\Delta_3}(P_3)\rangle = R^{\frac{5-d}{2}}\langle {\cal O}_{\Delta_1}(P_1){\cal O}_{\Delta_2}(P_2){\cal O}_{\Delta_3}(P_3)\rangle_{R=1}.
\end{equation}
\label{fn:C-scale}} We will also make the factors of $R$ manifest in the normalization \eqref{eq:norm} by denoting $N_\Delta = R^{\Delta-\frac{(d-1)}{2}}\hat{N}_\Delta$, with $\hat{N}_{\Delta}$ independent of $R$. 

The renormalized three-point function can then be written as
\begin{equation}
\label{eq:ren-three-pt}
    \prod_{i}\frac{1}{N_{\Delta_i}}\langle{\cal O}_1{\cal O}_2{\cal O}_3\rangle = \left(\prod_i\dfrac{1}{\hat{N}_{\Delta_i}}\right) \dfrac{R^{\frac{3(d-1)}{2}+\frac{5-d}{2}-\sum_i\Delta_i}\hat{C}_{123}}{(-P_1\cdot P_2+i\epsilon)^{\frac{\alpha_{12}}{2}}(-P_1\cdot P_3+i\epsilon)^{\frac{\alpha_{13}}{2}}(-P_2\cdot P_3+i\epsilon)^{\frac{\alpha_{23}}{2}}},
\end{equation}
and the next step is to evaluate its large $R$ expansion. Naively one could consider expanding each of the two-point structures individually using \eqref{eq:2pt-expansion}, but as we discuss in Appendix \ref{sec:three-point-naive} this approach suffers from issues due to the multiplication of distributions. Furthermore, the scaling with $R$ of the 8 terms obtained in this way is summarized in Table \ref{tab:R-scaling}.
\begin{table}[h!]
\begin{center}
    \begin{tabular}{ |l|l|l|l| }
  \hline
  Three delta functions & \multicolumn{3}{|c|}{$R^{3\frac{1-d}{2} {+ \sigma}}$} \\
  \hline
  Two delta functions & $R^{\frac{1-d}{2}+\Delta_1-\Delta_2-\Delta_3{+ \sigma}}$ & $R^{\frac{1-d}{2}+\Delta_2-\Delta_1-\Delta_3 {+ \sigma}}$ & $R^{\frac{1-d}{2}+\Delta_3-\Delta_1-\Delta_2{+ \sigma}}$\\
  \hline
  One delta function & $R^{-\frac{1-d}{2}-2\Delta_3{+ \sigma}}$ & $R^{-\frac{1-d}{2}-2\Delta_2{+ \sigma}}$ & $R^{-\frac{1-d}{2}-2\Delta_1{+ \sigma}}$\\
  \hline
  No delta function & \multicolumn{3}{|c|}{$R^{3\frac{d-1}{2}-\Delta_1-\Delta_2-\Delta_3{+ \sigma}}$}\\
  \hline
\end{tabular}
\end{center}
 \caption{Powers of $R$ multiplying the various three-point structures appearing in the dimensionally reduced three-point functions. $\sigma$ is the scaling of the OPE coefficient with $R$. For the scalar 3-point vertex, $\sigma = \frac{5 - d}{2}$.}
  \label{tab:R-scaling}
  \end{table}
  We see that all these terms are suppressed for generic real positive dimensions. We now show that \eqref{eq:ren-three-pt} contains an additional distributional term at $O(1)$ in the large $R$ expansion which is proportional to the scalar celestial three-point amplitude.

We would like to instead treat the three factors in \eqref{eq:ren-three-pt} on the same footing and expand their product at large $R$. To this end, we employ the Mellin representation of the three two-point structures
\begin{equation}
    \begin{split}
        \prod_{i}\frac{1}{N_{\Delta_i}}\langle{\cal O}_1{\cal O}_2{\cal O}_3\rangle &= \left(\prod_i\dfrac{1}{\hat{N}_{\Delta_i}}\right) \dfrac{({-}i)^{\frac{\Delta_1+\Delta_2+\Delta_3}{2}}R^{d+1-\sum_i\Delta_i}\hat{C}_{123}}{\Gamma(\frac{\alpha_{12}}{2})\Gamma(\frac{\alpha_{13}}{2})\Gamma({\frac{\alpha_{23}}{2}})}  \prod_{i}\int_0^\infty d\omega_i \omega_i^{\alpha_{jk}/2-1}\\
        &\times e^{i\omega_3\left(-P_1\cdot P_2+i\epsilon\right)}e^{i\omega_2\left(-P_1\cdot P_3+i\epsilon\right)}e^{i\omega_1\left(-P_2\cdot P_3+i\epsilon\right)}, \quad  j \neq k \neq i \in \{1,2,3 \}.
    \end{split}
\end{equation}
Observe that in this representation, the dependence on $R$ is contained in the three exponents which can be combined prior to taking the large-$R$ limit. This approach evades the issues associated with multiplication of distributions. At large $R$ the exponents can be replaced by their expansion \eqref{eq:two-point-exp-app} and we obtain
\begin{equation}
    \begin{split}
        \prod_{i}\frac{1}{N_{\Delta_i}}\langle{\cal O}_1{\cal O}_2{\cal O}_3\rangle &= \left(\prod_i\dfrac{1}{\hat{N}_{\Delta_i}}\right) \dfrac{({-}i)^{\frac{\Delta_1+\Delta_2+\Delta_3}{2}}R^{d+1-\sum_i\Delta_i}\hat{C}_{123}}{\Gamma(\frac{\alpha_{12}}{2})\Gamma(\frac{\alpha_{13}}{2})\Gamma({\frac{\alpha_{23}}{2}})}  \prod_{i}\int_0^\infty d\omega_i \omega_i^{\alpha_{jk}/2-1}\\
        &\times e^{i\frac{\omega_3}{2}\eta_1\eta_2\left(|\Omega_{12}|^2-\frac{u_{12}^2}{R^2}+i\epsilon\right)}e^{i\frac{\omega_2}{2}\eta_1\eta_3\left(|\Omega_{13}|^2-\frac{u_{13}^2}{R^2}+i\epsilon\right)}e^{i\frac{\omega_1}{2}\eta_2\eta_3\left(|\Omega_{23}|^2-\frac{u_{23}^2}{R ^2}+i\epsilon\right)}.
    \end{split}
\end{equation}
Now rescaling $\omega_i\to 2\omega_i$ and afterwards changing variables according to 
\begin{equation}
    \omega_1 = \omega_2'\omega_3',\quad \omega_2=\omega_1'\omega_3',\quad \omega_3 = \omega_1'\omega_2',
\end{equation}
we obtain
\begin{equation}
    \begin{split}
        \prod_{i}\frac{1}{N_{\Delta_i}}\langle{\cal O}_1{\cal O}_2{\cal O}_3\rangle &= \left(\prod_i\dfrac{1}{\hat{N}_{\Delta_i}}\right) \dfrac{2^{\frac{\Delta_1+\Delta_2+\Delta_3}{2}+1}({-}i)^{\frac{\Delta_1+\Delta_2+\Delta_3}{2}}R^{d+1-\sum_i\Delta_i}\hat{C}_{123}}{\Gamma(\frac{\alpha_{12}}{2})\Gamma(\frac{\alpha_{13}}{2})\Gamma({\frac{\alpha_{23}}{2}})}  \prod_{i}\int_0^\infty d\omega_i \omega_i^{\Delta_i-1}\\
        &\times e^{i{\eta_1\omega_1\eta_2\omega_2}\left(|\Omega_{12}|^2-\frac{u_{12}^2}{R^2}+i\epsilon\right)}e^{i{\eta_1\omega_1\eta_3\omega_3}\left(|\Omega_{13}|^2-\frac{u_{13}^2}{R^2}+i\epsilon\right)}e^{i{\eta_2\omega_2\eta_3\omega_3}\left(|\Omega_{23}|^2-\frac{u_{23}^2}{R^2}+i\epsilon\right)}.
    \end{split}
\end{equation}
Using the parametrization \eqref{eq:round-momenta}, 
\begin{equation}
\sum_{i<j}\eta_i\omega_i\eta_j\omega_j|\Omega_{ij}|^2=-\left(\sum_{i}\eta_i\omega_i \hat{q}_i\right)^2
\end{equation}
 and rescaling $\omega_i\to R\omega_i$ we are finally left with
\begin{equation}
    \begin{split}
        \prod_{i}\frac{1}{N_{\Delta_i}}\langle{\cal O}_1{\cal O}_2{\cal O}_3\rangle &= \left(\prod_i\dfrac{1}{\hat{N}_{\Delta_i}}\right) \dfrac{2^{\frac{\Delta_1+\Delta_2+\Delta_3}{2}+1}({-}i)^{\frac{\Delta_1+\Delta_2+\Delta_3}{2}}\hat{C}_{123}}{\Gamma(\frac{\alpha_{12}}{2})\Gamma(\frac{\alpha_{13}}{2})\Gamma({\frac{\alpha_{23}}{2}})}  \prod_{i}\int_0^\infty d\omega_i \omega_i^{\Delta_i-1}\\
        &\times e^{-i{\eta_1\omega_1\eta_2\omega_2}u_{12}^2+i\eta_1\omega_1\eta_2\omega_2\epsilon}e^{-i{\eta_1\omega_1\eta_3\omega_3}u_{13}^2+i\eta_1\omega_1\eta_3\omega_3\epsilon}e^{-i{\eta_2\omega_2\eta_3\omega_3}u_{23}^2+i\eta_2\omega_2\eta_3\omega_3\epsilon}\\
        &\times R^{d+1}e^{-iR^2(\sum_i\eta_i\omega_i \hat{q}_i)^2}.
    \end{split}
\end{equation}
At this point we recall the delta function Gaussian identity whose proof we review in Appendix \ref{app:delta-thre-pt}
\begin{equation}
        \lim_{\epsilon\to 0}\dfrac{1}{(2\pi\epsilon)^{n/2}}e^{-\frac{1}{2\epsilon}x^2} = \delta^{(n)}(x),\quad x\in\mathbb{R}^n.
    \end{equation}
Setting $\epsilon\to -\frac{i}{2}\epsilon$ we obtain
\begin{equation}
        \lim_{\epsilon\to 0}\dfrac{1}{(-i\pi\epsilon)^{n/2}}e^{-\frac{i}{\epsilon}x^2} = \delta^{(n)}(x),\quad x\in\mathbb{R}^n.
    \end{equation}
 Identifying $\epsilon = \frac{1}{R^{ 2}}$  we can use this identity to take the $R\to \infty$ limit and therefore obtain
\begin{equation}
\label{eq:three-point-int}
    \begin{split}
       \lim_{R \rightarrow \infty} \prod_{i}\frac{1}{N_{\Delta_i}}\langle{\cal O}_1{\cal O}_2{\cal O}_3\rangle &= \left(\prod_i\dfrac{1}{\hat{N}_{\Delta_i}}\right) \dfrac{\pi^{\frac{d+1}{2}}2^{\frac{\Delta_1+\Delta_2+\Delta_3}{2}+1}({-}i)^{\frac{\Delta_1+\Delta_2+\Delta_3+d+1}{2}}\hat{C}_{123}}{\Gamma(\frac{\alpha_{12}}{2})\Gamma(\frac{\alpha_{13}}{2})\Gamma({\frac{\alpha_{23}}{2}})}  \prod_{i}\int_0^\infty d\omega_i \omega_i^{\Delta_i-1}\\
        &\times e^{-i{\eta_1\omega_1\eta_2\omega_2}u_{12}^2+i\eta_1\omega_1\eta_2\omega_2\epsilon}e^{-i{\eta_1\omega_1\eta_3\omega_3}u_{13}^2+i\eta_1\omega_1\eta_3\omega_3\epsilon}e^{-i{\eta_2\omega_2\eta_3\omega_3}u_{23}^2+i\eta_2\omega_2\eta_3\omega_3\epsilon}\\
        &\times \delta^{(d+1)}\left(\sum_i\eta_i\omega_i \hat{q}_i\right).
    \end{split}
\end{equation}
We already recognize the Mellin transform of the scalar three-point amplitude computed in Section \ref{sec:celestial-amplitudes} and it remains to show that the prefactors correctly combine to  $(2\pi)^{d+1}g$. As discussed in Section \ref{sec:preliminaries}, we can either evaluate the time-Mellin transforms of the expanded correlator, or shift $\Delta_i\to \Delta_i^{\rm CCFT}$ followed by the limit $u_i\to 0$.  We evaluate the $u_i$ integrals in Appendix \ref{app:3pt-time-mellin}. After some work, this first approach yields a result that only agrees with the expectation after evaluating the residue at $\sum_{i = 1}^3\Delta_i - d - 2$, see \eqref{eq:other result}. We will comment on potential sources of this discrepancy later. We will focus on the second approach which generates exactly the expected celestial three-point function upon substituting the OPE coefficient $\hat{C}_{123}$ for that of a $\phi^3$ bulk vertex (see for example \cite{freedman1999correlation}).

From \eqref{eq:new-prescription} and \eqref{eq:three-point-int} with $\Delta_i \rightarrow \Delta_i^{\rm CCFT}$ and $u_i = 0$, we expect
\begin{equation}
    \begin{split}
        \lim_{R \rightarrow \infty} \prod_{i}\frac{1}{N_{\Delta_i}}&\int_{-\infty}^\infty du_i (u_i+i\eta_i\epsilon)^{-\lambda_i}\langle{\cal O}_1{\cal O}_2{\cal O}_3\rangle \\
        &= \prod_{i=1}^n \left(\frac{i^{\Delta_i}\Gamma(\Delta_i)}{C^d_{\Delta_i}} \right)\dfrac{\pi^{\frac{d+1}{2}}2^{\frac{\Delta_1+\Delta_2+\Delta_3}{2}+1}({-}i)^{\frac{\Delta_1+\Delta_2+\Delta_3+d+1}{2}}\hat{C}_{123}}{\Gamma(\frac{\alpha_{12}}{2})\Gamma(\frac{\alpha_{13}}{2})\Gamma({\frac{\alpha_{23}}{2}})}\\
        &\times
     \prod_{i}\left(\int_0^\infty d\omega_i \omega_i^{\Delta_i-1} \right)\delta\left(\sum_i\eta_i\omega_i q_i\right)\bigg|_{\Delta_i\to \Delta_i^{\rm CCFT}}.
    \end{split}
\end{equation}
In the last line we recognize the CCFT$_{d-1}$ amplitude
\begin{equation}
    \begin{split}
       \lim_{R \rightarrow \infty} \prod_{i}\frac{1}{N_{\Delta_i}}\int_{-\infty}^\infty du_i (u_i+i\eta_i\epsilon)^{-\lambda_i}\langle{\cal O}_1{\cal O}_2{\cal O}_3\rangle &= {\cal N}\widetilde{A}_3\left(\Delta_1^{\rm CCFT},\Delta_2^{\rm CCFT},\Delta_3^{\rm CCFT}\right),
    \end{split}
\end{equation}
where we have defined
\begin{equation}
    {\cal N} \equiv \prod_{i=1}^n \left(\frac{i^{\Delta_i}\Gamma(\Delta_i)}{C_{\Delta_i}^{d}}\right) \frac{1}{g} \dfrac{\pi^{\frac{d+1}{2}}2^{\frac{\Delta_1+\Delta_2+\Delta_3}{2}+1}({-}i)^{\frac{\Delta_1+\Delta_2+\Delta_3+d+1}{2}}\hat{C}_{123}}{(2\pi)^{d+1}\Gamma(\frac{\alpha_{12}}{2})\Gamma(\frac{\alpha_{13}}{2})\Gamma({\frac{\alpha_{23}}{2}})}\bigg|_{\Delta_i\to \Delta_i^{\rm CCFT}}\,.
\end{equation}
The final step amounts to showing that ${\cal N} = 1$. Note that since ${\cal N}$ multiplies the celestial amplitude, which as we have already proved contains a $\delta(\sum_i \Delta_i^{\rm CCFT} - d -1)$, it should always be evaluated subject to the constraint $\sum_i \Delta_i^{\rm CCFT} = d+1$. 

On the one hand the normalization factors used with this prescription can be seen to be
\begin{equation}
    \prod_{i=1}^n \frac{i^{\Delta_i}\Gamma(\Delta_i)}{C_{\Delta_i}^{d}} = i^{\Delta_1+\Delta_2+\Delta_3}(2\pi^{d/2})^3 \prod_i\Gamma\left(\Delta_i-\frac{d}{2}+1\right),
\end{equation}
while on the other hand the OPE coefficient from \cite{freedman1999correlation} with our conventions for the normalization of the bulk-to-boundary propagator in AdS$_{d+1}$ is
\begin{equation}
    \hat{C}_{123} = g \frac{2^{\frac{\Delta_1+\Delta_2+\Delta_3}{2}-3}\Gamma\left(\frac{\alpha_{12}}{2} \right)\Gamma\left(\frac{\alpha_{13}}{2}\right)\Gamma\left(\frac{\alpha_{23}}{2}\right)\Gamma\left(\frac{\Delta_1+\Delta_2+\Delta_3 - d }{2} \right)}{2 \pi^d \Gamma\left(\Delta_1 - \frac{d}{2} + 1\right)\Gamma\left(\Delta_2 - \frac{d}{2} + 1\right)\Gamma\left(\Delta_3 - \frac{d}{2} + 1 \right)}.
\end{equation}
As such we observe that ${\cal N}$ simplifies considerably to
\begin{equation}
    \begin{split}
        {\cal N} &= \dfrac{(2\pi^{d/2})^3i^{\Delta_1+\Delta_2+\Delta_3}\pi^{\frac{d+1}{2}}2^{\frac{\Delta_1+\Delta_2+\Delta_3}{2}+1}({-}i)^{\frac{\Delta_1+\Delta_2+\Delta_3+d+1}{2}}}{(2\pi)^{d+1}}\\
        &\times \frac{2^{\frac{\Delta_1+\Delta_2+\Delta_3}{2}-3}\Gamma\left(\frac{\Delta_1+\Delta_2+\Delta_3 - d }{2} \right)}{2 \pi^d }\bigg|_{\Delta_i\to \Delta_i^{\rm CCFT}}\,.
    \end{split}
\end{equation}
We finally recall the constraint on CCFT dimensions implied by the celestial amplitude, that gives
\begin{equation}
    \begin{split}
        {\cal N} &=  \dfrac{(2\pi^{d/2})^3i^{d+1}\pi^{\frac{d+1}{2}}2^{\frac{d+1}{2}+1}({-}i)^{d+1}}{(2\pi)^{d+1}}\frac{2^{\frac{d+1}{2}-3}\Gamma\left(\frac{1}{2} \right)}{2 \pi^d }=1\,.
    \end{split}
\end{equation}
Consequently we recover the celestial amplitude
\begin{equation}
    \begin{split}
      \lim_{R \rightarrow \infty}  \prod_{i}\frac{1}{N_{\Delta_i}}\int_{-\infty}^\infty du_i (u_i+i\eta_i\epsilon)^{-\lambda_i}\langle{\cal O}_1{\cal O}_2{\cal O}_3\rangle &= \widetilde{A}_3\left(\Delta_1^{\rm CCFT},\Delta_2^{\rm CCFT},\Delta_3^{\rm CCFT}\right).
    \end{split}
\end{equation}

As discussed in detail in Section \ref{sec:preliminaries}, the Witten diagram analysis suggests that shifting $\Delta_i\to \Delta_i^{\rm CCFT}$ and taking $u_i\to 0$ should be equivalent to taking the time-Mellin transforms. This was explicitly verified for the two-point function in Appendix \ref{app:dim-reduction-ui->0}. For the three-point function, we show in Appendix \ref{app:3pt-time-mellin} that a direct computation of the time-Mellin transforms of \eqref{eq:three-point-int} leads to an additional trigonometric function in the normalization, namely
\begin{equation}
\label{eq:other result}
    \begin{split}
       \lim_{R \rightarrow \infty} \prod_{i}\frac{1}{N_{\Delta_i}}\int_{-\infty}^\infty du_i (u_i+i\eta_i\epsilon)^{-\lambda_i}\langle{\cal O}_1{\cal O}_2{\cal O}_3\rangle &= \dfrac{ i^{\sum_i \Delta_i - d - 1}}{2\sin \pi\frac{d+2-\sum_i\Delta_i}{2}}\\
       &\times \widetilde{A}_3\left(\Delta_1^{\rm CCFT},\Delta_2^{\rm CCFT},\Delta_3^{\rm CCFT}\right).
    \end{split}
\end{equation}
 While the prefactor can be removed by taking the residue at $\sum_i \Delta_i = d+2$ (up to a factor of 2), we were not able to identify the precise source of the mismatch. One can show that the integrand on the LHS of \eqref{eq:other result} contains branch cuts in the complex $\sum_i u_i \omega_i$. Quite interestingly, as explained in Appendix \ref{app:3pt-time-mellin}, one may obtain precisely the correct celestial 3-point function by considering a different integration path in the uplift of this plane to its universal cover. We leave a better understanding of this to future work.

\section{Discussion}\label{sec:discussion}

In this paper we  verified \eqref{eq:exp} which relates renormalized correlation functions in Lorentzian CFT$_d$ to celestial amplitudes in CCFT$_{d-1}$. For simplicity, we have focused on scalar two- and three-point correlators. We showed that celestial two-point functions are exactly reproduced from two-point correlators in a one higher dimensional unitary CFT by \eqref{eq:exp}. We presented two derivations of this relation. The first relied on applying the delta function identity \eqref{eq:2pt-expansion} to the CFT$_d$ two-point function followed by direct evaluation of the time-Mellin integrals in \eqref{eq:exp}. The second relied on \eqref{eq:new-prescription}, namely shifting the CFT$_d$ dimensions and evaluating the correlator on the Lorenztian cylinder on time slices separated by exactly $\pi$ in global time. The correct celestial two-point function \eqref{eq:two-point-final} including the normalization was obtained in both cases. 

In the case of three-point functions, we found that \eqref{eq:new-prescription} precisely reproduces the expected Euclidean celestial three-point functions of scalars. The derivation crucially relied on a careful expansion of the corresponding Lorentzian CFT$_d$ correlator by employing the delta function representation \eqref{app:delta-thre-pt}. On the other hand, direct evaluation of the time Mellin integrals in \eqref{eq:exp} after applying the same expansion led to a mismatch by a trigonometric function involving the sum of the dimensions of the CFT$_d$ operators. In both cases, the results are manifestly Poincar\'e invariant. 

Already at the level of two-point functions, our calculation revealed a series of interesting take-home messages. We have shown that CFT$_d$ two-point functions develop transverse delta function singularities in certain kinematic configurations. The normalization \eqref{eq:normz} found in \cite{PipolodeGioia:2022exe} to relate celestial amplitudes to AdS-Witten diagrams projects the correlators onto their singular component in the limit $R \rightarrow \infty$.  Furthermore, the integral transforms in \eqref{eq:exp} lead to the celestial two-point functions obtained by Mellin transforming the on-shell two-point ``scattering amplitude'' (or on-shell propagator). Prior to performing these integrals, the correlators are non-vanishing irrespective of whether the insertions are incoming or outgoing. The integral transforms precisely set to 0 the configurations where both insertions are incoming or outgoing. Keeping careful track of the $i\epsilon$ prescriptions appearing both in the integral transforms and the CFT$_d$ Lorentzian correlator was important to obtain the correct normalization. 

In the case of the three-point functions, a careful treatment of the expansion was needed in order to obtain the term which precisely coincides with the celestial three-point function of scalars after appropriately normalizing the CFT$_d$ operators. Furthermore, for generic dimensions above the CFT$_d$ unitarity bounds, other terms in the bulk point expansion were shown to be suppressed. The correct normalization of the celestial three-point function was shown to be reproduced by applying \eqref{eq:new-prescription}. A different integration path in the uplift of the complex $\sum_i \omega_i u_i$ plane to its universal cover had to be considered in order to obtain the correct normalization by directly evaluating the $u_i$ integrals in \eqref{eq:exp}.

{\bf Lorentzian CCFT.} Three-point scattering amplitudes are typically computed in bulk split signature. For a 4-dimensional bulk, the celestial sphere becomes a Lorentzian torus \cite{Atanasov:2021oyu}. One may expect the more commonly encountered Lorentzian formulas to be obtained from \eqref{eq:ccft-res-d} by simply using \eqref{eq:delta-fct-mom}, or its analog \eqref{eq:delta-d3} in $d = 3$. To be explicit, from \eqref{eq:ccft-res} we can immediately get Lorentzian celestial amplitudes of the form 
\be 
\begin{split}
 \tilde{A}^{\pm 1,\mp 1,\mp 1} &= \frac{1}{4}(2\pi)^4 g B(\Delta_{\rm CCFT}^2-1,\Delta_{\rm CCFT}^3-1)\delta\left(\sum_{i=1}^3\Delta_{\rm CCFT}^i-4\right)\delta(|z_{12}|^2)\delta(|z_{13}|^2)\\
 &= \frac{1}{4z_{12}z_{13}}(2\pi)^4 g B(\Delta_{\rm CCFT}^2-1,\Delta_{\rm CCFT}^3-1)\delta\left(\sum_{i=1}^3\Delta_{\rm CCFT}^i-4\right)\delta(\bz_{12})\delta(\bz_{13}).
 \end{split}
\ee
Note however that the standard dependence on the left-moving variables (see eg. \cite{Stieberger:2018edy}) cannot be resolved here because our starting point is a configuration with $z_{ij} = \bz_{ij} = 0$ (as opposed to one where $z_{ij} \neq 0$, $\bz_{ij} = 0$ or vice-versa). Furthermore, the normalization we obtain in this way will generally differ from the one where one starts with a split signature amplitude. This can be seen by comparing the rewriting of the momentum conserving delta function \eqref{eq:3-pt-delta} with that in for example eq. (2.5) of \cite{Stieberger:2018edy}. It will be interesting to better understand the analytic continuation from the celestial sphere to the Lorentzian torus which appears to not be unique at least in the case of three-point functions which are highly constrained by momentum conservation. It should be straightforward to generalize our analysis to other integer spins. 

{\bf Carrollian correlators.} Carrollian 2-, 3- and 4-point amplitudes were recently computed in \cite{Mason:2023mti} by evaluating the Fourier transforms of MHV scattering amplitudes in bulk $(2,2)$ signature. We compare and contrast the results obtained therein with what we found. In the case of two-point functions, our bulk-point limit yields a generalization of the formula (4.6) in \cite{Mason:2023mti}, which may be obtained from our \eqref{eq:C-tilde} with $d = 3$ by taking the limit $\Delta \rightarrow 1$. In the case of three-point functions, the Carrollian correlators obtained in \cite{Mason:2023mti} are clearly distinct from our three-point correlators \eqref{eq:three-point-int} where the time and angular dependence is factorized. This difference may be traced back to the fact that our prescription naturally computes Euclidean rather than Lorentzian amplitudes. We were not able to derive the Carrollian three-point functions of \cite{Mason:2023mti} directly from CFT$_3$ correlators. Note also that our derivation of celestial amplitudes, or Carrollian correlators prior to taking the integral transforms in \eqref{eq:exp}, is completely independent from the one in \cite{Bagchi:2023fbj}, who resorted to the relation between conformal correlators and AdS-Witten diagrams which by definition \cite{PipolodeGioia:2022exe} computes Carrollian correlators in the flat space limit. We emphasize that our prescription provides a bulk-independent way of deriving celestial and Carrollian correlators from a limit of a CFT correlator in one higher dimension. The latter clearly contains more structure a-priori. Already at the level of two-point functions, we see that additional input from the extrapolate dictionary (namely setting $\Delta = 1$ in the case of 4 bulk dimensions) is needed in order to obtain a Carrollian correlator. It would be interesting to better understand this relation for three- and higher-point functions and well as higher-dimensional theories. More generally, the dimensional reduction of CFT employed here may provide new insights into the peculiar structures appearing in celestial and Carrollian correlators, such as multi-particle and subleading collinear or OPE limits. 

{\bf 3-point correlator.} In our derivation of the three-point scalar CCFT$_{d-1}$ amplitude from CFT$_d$ we encountered a puzzle. From the general arguments outlined in \cite{PipolodeGioia:2022exe, de2023celestial},\footnote{We repeat the argument here for clarity, since such arguments have been previously used to compute Carrollian correlators in eg. \cite{Bagchi:2023fbj}. A scalar three-point CFT$_d$ correlator corresponding to a contact interaction in AdS$_{d+1}$ is given by
\be 
C_3 = \int_{AdS_{d+1}} d^{d+1}x K_{\Delta_1}(p_1;x)  K_{\Delta_2}(p_2;x)  K_{\Delta_3}(p_3;x), 
\ee
where $p_i$ are points on the boundary. In the approximate bulk point configuration, and after integrating over the infinitesimal strips, the bulk-to-boundary propagators reduce to conformal primary wavefunctions, while the $i\epsilon$ prescription inherited from the propagators corresponds to incoming or outgoing configurations
\be 
\begin{split}
C_3 &\rightarrow N\int_{\mathbb{R}^{d,1}} d^{d+1}x \varphi_{\Delta_1 + \lambda_1 - 1}(x;\hat{q}_1)\varphi_{\Delta_2 + \lambda_2 - 1}(x;\hat{q}_2)\varphi_{\Delta_3 + \lambda_3 - 1}(x;\hat{q}_3) \\
&\propto N \int d^{d+1}x \prod_{i = 1}^3 \int_0^{\infty} d\omega_i \omega_i^{\Delta_i + \lambda_i - 2} e^{i\sum_j \eta_j \omega_j \hat{q}_j \cdot x} \propto N \prod_{i = 1}^3 \int_0^{\infty} d\omega_i \omega_i^{\Delta_i + \lambda_i - 2} \delta^{(d+1)}\left(\sum_{j} \eta_j \omega_j \hat{q}_j\cdot x \right).
\end{split}
\ee
This should coincide exactly with the celestial amplitudes computed in Section \ref{sec:celestial-amplitudes} keeping careful track of the normalization (as we did). }
we expected the celestial three-point function to be exactly reproduced by applying either \eqref{eq:exp} or \eqref{eq:new-prescription}.  While the latter formula was shown to lead to the expected result including the correct normalization, the former was found to only give agreement after further evaluating a residue at $\sum_{i}\Delta_i  = d + 2.$ We were not able to identify the precise source of this mismatch. Regarding the CFT correlator as computed through an AdS Witten diagram, it may be that the AdS integrals simply don't commute with the integrals over the strips separated by $\pi$ in global time on the Lorentzian cylinder. On the other hand, the simple strucutre of the mismatch suggests rather that a subtlety has been overlooked in our computation in Appendix \ref{app:3pt-time-mellin}. Perhaps a more careful consideration of the $i\epsilon$ prescriptions involved in the definition of the Lorentzian three-point function will generate the missing $2 \sin \pi \frac{d + 2 - \sum_i \Delta_i}{2}$ in our result \eqref{eq:other result}. In fact, we have already found in Appendix \ref{app:3pt-time-mellin} an integration path in the uplift of the complex $v_3 = \sum_i \omega_i u_i$ plane to its universal cover that yields precisely the correct celestial amplitude. It remains to explain this choice and understand its physical meaning.

{\bf Higher-point functions.} The analysis we performed here should serve as a warm-up for the more interesting case of higher-point functions. As first pointed out in \cite{maldacena2017looking} and analyzed more recently in a different setup in \cite{jain2023s}, Lorentzian four-point functions develop bulk-point singularities at real cross-ratios. This can be seen by analyzing the solutions to the conformal Casimir equations in the limit $Z \rightarrow \bar Z$, which develop singularities of the form
\begin{equation}
    G_{\Delta,\ell}(Z,\bar Z) = \dfrac{1}{(Z-\bar Z)^{d-3}}f_{\Delta,\ell}(Z,\bar Z),\quad \text{as $Z\to \bar Z$}   , 
\end{equation}
where $(Z, \bar{Z})$ are conformal cross-ratios and where $f_{\Delta,\ell}(Z,\bar Z)$ is regular in the bulk point limit. That these singularities should be related to similar singularities in four-point celestial amplitudes due to momentum conservation was already pointed out in \cite{Lam:2017ofc} but we can also motivate it from our analysis. The cross-ratios $(Z,\bar Z)$ in the CFT$_d$ are functions of the four insertion points $P(\tau_i,\eta_i\Omega_i)$ and therefore admit an expansion about the flat space limit configuration $\tau_i = \eta_i\frac{\pi}{2}+\frac{u_i}{R}$, with the property that the cross ratios expand at large $R$ as $(Z,\bar Z) = (z,\bz) + O(R^{-2})$ where $(z,\bz)$ are the CCFT$_{d-1}$ cross ratios formed from the points $\Omega_i$ on $S^{d-1}$. It follows that we can apply the Dirac delta identity \eqref{eq:dirac-delta-expansion}. In that scenario, the AdS bulk-point singularity $(Z-\bar Z)^{3-d}$ becomes the $\delta(z-\bz)$ present in all 4-point celestial amplitudes under dimensional reduction. It would be interesting to carefully analyze the behavior of conformal blocks on the Lorentzian cylinder in the limit considered here. In the cases of interest (such as 2d CCFT), this will be complicated by the fact that no closed formulas for conformal blocks in odd dimensions are known. In \cite{maldacena2017looking} it was argued that the bulk-point singularities are erased by quantum gravity effects for $(d+2)$-point functions in $d$-dimensional CFT. This is in line with the fact that low-point amplitudes in CCFT$_{d-1}$ will always have singularities due to bulk momentum conservation, while for higher-point functions the Mellin integrals with respect to the external energies may saturate all of the delta functions from momentum conservation. It will also be interesting to understand whether conformal blocks in the bulk point limit can be related to integrated products of the distributional three-point functions obtained here. This may be useful in finding a decomposition of celestial higher-point functions in terms of a more suitable (yet potentially singular) basis of solutions to the conformal Casimir equation. 

{\bf Insights into holographic correlators from celestial CFT.} We note that the discrepancy encountered in our computation of the scalar celestial three-point function using the two flat space prescriptions disappears for certain integer linear combinations of the CFT$_d$ dimensions. Furthermore, in order for the flat space result to dominate in the bulk-point limit (ie. for the correlators to be consistent with the existence of a local bulk), the dimensions of the CFT$_d$ operators need to satisfy the unitarity bounds. It will be interesting to see if properties of celestial CFT could place any further constraints on holographic CFT in higher dimensions. For example, loop corrections will modify the three-point coefficient of the celestial amplitude, which should in turn be related to the anomalous dimensions acquired by operators in the higher dimensional CFT.  We leave these interesting avenues of research to the future.


\section*{Acknowledgements}

We thank Pinaki Banerjee, Connor Behan, Geoffrey Compere, Laurent Freidel, Shiraz Minwalla, Pedro Vieira and Yifei Zhao for discussions, as well as Pedro de Ornelas Sampaio Macedo for collaboration in the very early stages of the project. A.R. was supported by the European Commission through a Marie Sklodowska-Curie fellowship at the University of Amsterdam, grant agreement 101063234 and would like to acknowledge the Erwin Schr\"{o}dinger Institute for hospitality during the last stage of this project. L. G. would like to acknowledge the support provided by FAPESP Foundation through the grant 2023/04415-2.

\appendix

\section{Delta function identities}
\label{app:A}

In this appendix we prove two delta function identities that will play important roles in extracting celestial two- and three-point functions from Lorentzian conformal correlators. 

\subsection{Two-point function}

We start by proving the identity
\begin{equation}
\label{eq:delta-fct}
\begin{split}
  \left(\dfrac{\epsilon}{\epsilon^2+z^2}\right)^\Delta &=\left[\dfrac{\pi^{n/2}\Gamma\left(\Delta-\frac{n}{2}\right)}{\Gamma(\Delta)\epsilon^{\Delta-n}}\delta^{(n)}(z)+O(\epsilon^{n-\Delta+ 2})\right] +\left[\dfrac{\epsilon^\Delta}{z^{2\Delta}}+O(\epsilon^{\Delta+2})\right], \quad z\in\mathbb{R}^n.
    \end{split}
    \end{equation}
    
    Let $k \in \mathbb{R}^n$ and consider the Fourier transform\footnote{Alternatively it is possible to work in position space and integrate the LHS of \eqref{eq:delta-fct} against a test function $\phi\in C^\infty_0(\mathbb{R}^n)$ decomposing the integral over regions $|z|<\Lambda$ and $\Lambda<|z|$. Then the first region generates the delta function series and the second generates the power-law series. Moreover this method reveals that the $\frac{1}{|z|^{2\Delta+2m}}$ terms should be interpreted in the principal value sense.}
    \be 
    \label{eq:a2}
    \begin{split}
    \int d^nz e^{-i k\cdot z}   \left(\dfrac{\epsilon}{\epsilon^2+z^2}\right)^\Delta &=    \frac{1}{\Gamma(\Delta)}\int d^nz e^{-i k\cdot z} \int_0^{\infty} d\alpha \alpha^{\Delta - 1} e^{-\alpha \left(\epsilon + \frac{z^2}{\epsilon} \right)} \\
    &= \frac{1}{\Gamma(\Delta)} \int_0^{\infty}
 d\alpha \alpha^{\Delta - 1} e^{-\alpha \epsilon} e^{- \frac{\epsilon}{4 \alpha} k^2} \int d^nz e^{-\frac{\alpha}{\epsilon}\left(z + i \frac{\epsilon}{2\alpha} k \right)^2} \\
 & = \frac{1}{\Gamma(\Delta)} \int_0^{\infty}
 d\alpha \alpha^{\Delta - 1} e^{-\alpha \epsilon} e^{- \frac{\epsilon}{4 \alpha} k^2} (\pi \epsilon)^{n/2} \alpha^{-n/2}.
 \end{split}
     \ee
     In order to evaluate the $\epsilon \rightarrow 0$ limit, we use the method of regions \cite{Jantzen:2011nz,Smirnov:2001in} and decompose the integral over $\alpha$ into small- and large-$\alpha$ regimes:
     \be 
     \label{eq:a3}
     \begin{split}
  I &\equiv    \int_0^{\infty}
 d\alpha \alpha^{\Delta - 1 - n/2} e^{-\alpha \epsilon} e^{- \frac{\epsilon}{4 \alpha} k^2}  =    \int_{\Lambda}^{\infty}
 d\alpha \alpha^{\Delta - 1- n/2} e^{-\alpha \epsilon} e^{- \frac{\epsilon}{4 \alpha} k^2} \\
 & +    \int_0^{\Lambda}
 d\alpha \alpha^{\Delta - 1 - n/2} e^{-\alpha \epsilon} e^{- \frac{\epsilon}{4 \alpha} k^2}
  \equiv I_{>\Lambda} + I_{< \Lambda}.
 \end{split}
     \ee 
The advantage of this decomposition is that only one of the exponentials will be dominating for each regime in the $\epsilon \rightarrow 0$ limit, namely 
\be 
\label{regions}
\begin{split}
I_{> \Lambda} &= \Gamma\left(\Delta - \frac{n}{2}\right) \epsilon^{n/2 - \Delta} + O( \epsilon^{n/2 - \Delta + 2}) + \cdots, \quad {{\rm Re}\left(\Delta - \frac{n}{2}\right) > 0,} \\
I_{< \Lambda} &= - \int_{\infty}^{1/\Lambda} d\alpha \alpha^{-2} \alpha^{1 + n/2 - \Delta} e^{-\epsilon/\alpha} e^{-\frac{\epsilon}{4} \alpha k^2} = \Gamma\left(\frac{n}{2} - \Delta\right)\left(\frac{\epsilon k^2}{4} \right)^{\Delta - n/2} + O((\epsilon k^2)^{\Delta  - n/2 + 2}) + \cdots,
\end{split}
\ee
where $\cdots$ denote $\Lambda$-dependent terms that cancel upon taking the sum (which is manifestly independent on $\Lambda$). Note that $I_{>\Lambda}$ converges for ${\rm Re}\Delta > \frac{n}{2}$, while $I_{<\Lambda}$ converges for ${\rm Re}\Delta < \frac{n}{2}$. However, we can extend the regime of validity of $I_{<\Lambda}$ to  Re$\Delta > \frac{n}{2}$ by introducing a regulator and analytically continuing the second integral in \eqref{regions} to ${\rm Re}\left(\Delta -\frac{n}{2}\right) >0$. In fact the poles appearing on the RHS of this equation when $\Delta - \frac{n}{2} < 0$ will cancel in the final result. We hence restrict Re$\left(\Delta - \frac{n}{2}\right) > 0$. We can now evaluate the inverse Fourier transform of the expansions \eqref{regions},
\be 
\begin{split}
\frac{1}{(2\pi)^n}& \int d^nk e^{ik\cdot z} I_{>\Lambda} = \Gamma\left( \Delta - \frac{n}{2} \right) \epsilon^{n/2 - \Delta} \delta^{(n)}(z) + O( \epsilon^{n/2 - \Delta + 2}) + \cdots, \\
\frac{1}{(2\pi)^n}& \int d^nk e^{ik\cdot z} I_{<\Lambda} =  \Gamma\left(\frac{n}{2} - \Delta\right) \frac{1}{(2\pi)^n} \int d^nk e^{ik\cdot z} \left[\left(\frac{\epsilon k^2}{4} \right)^{\Delta - n/2} + O(\epsilon^{\Delta  - n/2 + 2}) \right] + \cdots\\
&=  \Gamma\left(\frac{n}{2} - \Delta\right) \frac{1}{(2\pi)^n} \int  d\Omega_{n-1} \int dk k^{n-1} e^{i k z \cos \theta} \left(\frac{\epsilon}{4}\right)^{\Delta - n/2} k^{2\Delta - n} + O(\epsilon^{\Delta - n/2 + 2}) + \cdots \\
& = (2\pi)^{-n} \Gamma\left(\frac{n}{2} - \Delta\right) \Gamma(2\Delta) 2^{n - 2\Delta} \epsilon^{\Delta - n/2} \int d\Omega_{n-1} (-i z\cos \theta)^{-2\Delta}+ O(\epsilon^{\Delta - n/2 + 2}) + \cdots \\
&=(2\pi)^{-n}  \Gamma(2\Delta) 2^{n - 2\Delta} \epsilon^{\Delta - n/2} z^{-2\Delta}\Gamma(1/2 - \Delta) 2\pi^{(n-1)/2} \cos \pi \Delta + O(\epsilon^{\Delta - n/2 + 2}) + \cdots .
\end{split}
\ee
Using the Legendre duplication formula, and multiplying by the factor we dropped in going from \eqref{eq:a2} to \eqref{eq:a3}, we recover \eqref{eq:delta-fct}. Note that the second term in \eqref{eq:delta-fct} could have been more easily obtained as the leading term in a Taylor expansion around $\epsilon = 0$. 

We will now use this formula to evaluate
\be 
\lim_{\beta \rightarrow 0} \left( \frac{\beta}{-\eta_1\eta_2 ({\zeta} \beta)^2 +\eta_1\eta_2 z^2 + i\epsilon} \right)^{\Delta} = \lim_{\beta \rightarrow 0} e^{-i\pi \Delta \theta(-\eta_1\eta_2)} \left(\frac{\beta}{z^2 + (i\zeta\beta)^2}\right)^{\Delta}, \quad \zeta = \pm 1.
\ee 
Here we take the branch cut of $(y + i\epsilon)^{\Delta}$ to be in the lower half plane (and upper half plane for $\epsilon \rightarrow -\epsilon$) which implies that 
\be 
(-y \pm i\epsilon)^{\Delta} = e^{\pm i\pi \Delta} |y|^\Delta, \quad y>0.
\ee
We can now follow through the same calculation above replacing $\beta \rightarrow i \zeta \beta$ to obtain  for ${\rm Re}\left(\Delta - \frac{n}{2} \right) > 0$
\be 
\label{eq:exp-epsilon}
\begin{split}
\lim_{\beta \rightarrow 0} &\left( \frac{\beta}{-\eta_1\eta_2({\zeta\beta})^2 +\eta_1\eta_2 z^2 + i\epsilon}\right)^{\Delta} = e^{-i\pi \Delta \theta(-\eta_1\eta_2)} (-i\zeta)^{\Delta}\\
&\times \left[\left(\dfrac{\pi^{n/2}\Gamma\left(\Delta-\frac{n}{2}\right)}{\Gamma(\Delta)(i\zeta\beta)^{\Delta-n}}\delta^{(n)}(z)+O(\beta^{n-\Delta+2})\right)+ \left(\dfrac{(i\zeta\beta)^{\Delta}}{z^{2\Delta}}+O(\beta^{\Delta+2})\right)\right], \quad z\in\mathbb{R}^n.
\end{split}
\ee

Now we note that
\be 
\frac{1}{(-P_1\cdot P_2 + i{\eta_1\zeta}\epsilon)^{\Delta}} = \frac{1}{\left(\cosh(\tau_1 - \tau_2) - \Omega_1 \cdot \Omega_2 + i{\eta_1\zeta}\epsilon\right)^{\Delta}}.
\ee
Analytically continuing to Lorentzian signature $\tau = i t$, and setting $t_1 = \eta_1 \frac{\pi}{2} + \frac{u_1}{R}$ and $t_2 = \eta_2 \frac{\pi}{2} + \frac{u_2}{R}$ with $\Omega_1$ and $\Omega_2$ antipodally related provided that $\eta_1\eta_2 = -1$ \cite{PipolodeGioia:2022exe,de2023celestial}, we find as $R \rightarrow \infty$
\be 
\label{eq:two-point-exp-app}
\frac{1}{(-P_1\cdot P_2 + i{\eta_1\zeta}\epsilon)^{\Delta}} = e^{-i\pi{\eta_1\zeta} \Delta \theta(-\eta_1\eta_2)}\left(\frac{2}{(\Omega_1 -  \Omega_2)^2 + \frac{(i(u_1-u_2) + \epsilon)^2}{R^2}}\right)^{\Delta} + O\left(R^{-4} \right).
\ee
At large-$R$ the two-point function develops a branch cut at $u_{12} = 0$ and one needs to specify an additional $i\epsilon$ prescription to the one appearing on the LHS of \eqref{eq:two-point-exp-app} which instead tells us how to cross the branch cut in the Lorentzian correlator at finite $|\Omega_{12}|, t_{12}$. A priori these $i\epsilon$ prescriptions may be independently specified.  Another way to see this is to allow for a large-$R$ expansion of the $\epsilon$ appearing in the analytic continuation of the Euclidean time
\be 
\tau = i t + \epsilon, \quad \epsilon = \epsilon^{(0)} + \frac{1}{R} \epsilon^{(1)} + O(R^{-2}).
\ee
In \eqref{eq:two-point-exp-app} we have chosen $\epsilon^{(0)} = \epsilon^{(1)} = \epsilon$.  Furthermore, by considering the different in-out configurations for the two points, one can show that the $i\eta_1 \zeta \epsilon$ prescription on the LHS of \eqref{eq:two-point-exp-app} computes the Wightman functions corresponding to $\zeta = \pm 1$ for either choice of $\eta_1 = \pm 1$.
Eq. \eqref{eq:2pt-expansion} then follows immediately from \eqref{eq:exp-epsilon} and \eqref{eq:two-point-exp-app}. 

\subsection{Three-point function}
\label{app:delta-thre-pt}

The following delta function identity will be useful in the analysis of three-point functions
\be 
\label{eq:delta-three-pt}
\begin{split}
\lim_{\epsilon \rightarrow 0} \frac{1}{(2\pi)^{n/2} \epsilon^n} e^{-\frac{1}{2}\frac{x^2}{\epsilon^2}} = \delta^{(n)}(x), \quad x \in \mathbb{R}^n.
\end{split}
\ee
To prove \eqref{eq:delta-three-pt}, we let
\be 
f_{\epsilon}(x) \equiv \frac{1}{(2\pi)^{n/2} \epsilon^n} e^{-\frac{1}{2}\frac{x^2}{\epsilon^2}}
\ee
and consider $\phi \in C_0^{\infty}(\mathbb{R}^n)$. We now evaluate the distributional action of $f_{\epsilon}$ on $\phi$
\be
\langle f_{\epsilon}, \phi \rangle = \int_{\mathbb{R}^n} d^nx f_{\epsilon}(x) \phi(x) = \int_{\mathbb{R}^n} d^n y f_{1}(y) \phi(\epsilon y),
\ee
where in the second equality we changed variables $x = \epsilon y$. Since $\phi$ has compact support, the integrand obeys
\be 
\left|f_1(y) \phi(\epsilon y) \right| \leq f_1(y) {\rm max} \phi.
\ee
Since the upper bound does not depend on $\epsilon$ for fixed $y$, the Lebesgue dominated convergence
theorem implies that we can take the $\epsilon \rightarrow 0$ limit inside the integral and therefore
\be 
\lim_{\epsilon \rightarrow 0} \langle f_{\epsilon}, \phi \rangle = \phi(0) = \langle\delta, \phi \rangle,
\ee
from which \eqref{eq:delta-three-pt} follows. 

\section{Dimensional reduction of CFT$_d$ two-point function}
\label{app:dim-red}

In this appendix we detail the steps leading to the celestial 2-point function \eqref{eq:two-point-final} by applying the prescription outlined in Section \ref{sec:preliminaries}. In Appendix \ref{app:direct} we apply the integrals \eqref{eq:exp} to the Lorentzian CFT$_d$ correlators expanded in the time strips, while in appendix \ref{app:dim-reduction-ui->0} we instead shift the dimensions and set $u_i \rightarrow 0$. As we will see, the two procedures lead to the same result \eqref{eq:two-point-final} which precisely agrees with the CCFT$_{d-1}$ two-point function.

\subsection{Integrating over the strips}
\label{app:direct}

In this section we compute the integral transforms of translationally invariant functions of time $f(u_1 - u_2)$
\begin{equation}
\label{eq:time-Mellin-integral}
    {\cal I}_{\eta_1\eta_2}^{\lambda_1\lambda_2}[f] \equiv \int_{-\infty}^{\infty} du_1 (u_1+i\eta_1\epsilon)^{-\lambda_1}\int_{-\infty}^\infty du_2 (u_2+i\eta_2\epsilon)^{-\lambda_2} f(u_1-u_2).
\end{equation}
It is convenient to change coordinates to  $(x,y)=(u_1-u_2,u_2)$, in which case \eqref{eq:time-Mellin-integral} becomes
\begin{equation}
    \begin{split}
        {\cal I}_{\eta_1\eta_2}^{\lambda_1\lambda_2}[f] 
        &=\int_{-\infty}^{\infty} \int_{-\infty}^\infty dx dy (x+y+i\eta_1\epsilon)^{-\lambda_1}  (y+i\eta_2\epsilon)^{-\lambda_2} f(x) \\
        &=\int_{-\infty}^{\infty} \int_{-\infty}^\infty dx dy (x+i\eta_1\epsilon)^{-\lambda_1}\left(1+\frac{y}{x+i\eta_1\epsilon}\right)^{-\lambda_1}(y+i\eta_2\epsilon)^{-\lambda_2} f(x) \\
        &=\int_{-\infty}^{\infty} \int_{-\infty}^\infty dx dy (x+i\eta_1\epsilon)^{-\lambda_1}\left(1+\frac{y}{x}-\frac{iy\eta_1\epsilon}{x^2}\right)^{-\lambda_1}(y+i\eta_2\epsilon)^{-\lambda_2} f(x) \\
        &=\int_{-\infty}^{\infty} \int_{-\infty}^\infty dx dy |x|(x+i\eta_1\epsilon)^{-\lambda_1}\left(1+y-\frac{iy\eta_1\epsilon}{x}\right)^{-\lambda_1}(xy+i\eta_2\epsilon)^{-\lambda_2} f(x).
    \end{split}
\end{equation}
Note that in the second line we have factored out $x$ taking the $i\epsilon$ prescription into account and in the third line we have expanded in small $\epsilon$. In the final line we have transformed $ y \to xy$. To finish decoupling the integrals we notice that
\begin{equation}
    \begin{split}
        xy+i\eta_2\epsilon = (x+i\eta_2\epsilon) \left(\dfrac{xy+i\eta_2\epsilon}{x+i\eta_2\epsilon}\right) =(x+i\eta_2\epsilon)\left(y-i\eta_2(y-1)\frac{\epsilon}{x}+O(\epsilon^2)\right).
    \end{split}
\end{equation}
As a result we have
\begin{equation}
    \begin{split}
        {\cal I}_{\eta_1\eta_2}^{\lambda_1\lambda_2}[f] &=\int_{-\infty}^{\infty} \int_{-\infty}^\infty dx dy |x| (x+i\eta_1\epsilon)^{-\lambda_1}(x+i\eta_2\epsilon)^{-\lambda_2}\left(1+y-\frac{iy\eta_1\epsilon}{x}\right)^{-\lambda_1}\\
        &\times \left(y-i\eta_2(y-1)\frac{\epsilon}{x}\right)^{-\lambda_2} f(x).
    \end{split}
\end{equation}
Since $\epsilon$ is infinitesimal only the sign of what multiplies it matters. As such, we can define the following function, where $\sigma \equiv \operatorname{sgn}(x)$

\begin{equation}
    I_{\eta_1\eta_2}^\sigma(\lambda_1,\lambda_2) \equiv \int_{-\infty}^{\infty}dy \left(1+y-i\operatorname{sgn}(y)\sigma \eta_1\epsilon\right)^{-\lambda_1} \left(y -i\operatorname{sgn}(y-1)\sigma\eta_2\epsilon\right)^{-\lambda_2},
\end{equation}
so that 
\begin{equation}
    {\cal I}^{\lambda_1\lambda_2}_{\eta_1\eta_2}[f] = \int_{-\infty}^\infty dx |x|(x+i\eta_1\epsilon)^{-\lambda_1}(x+i\eta_2\epsilon)^{-\lambda_2}I^{\sigma}_{\eta_1\eta_2}(\lambda_1,\lambda_2)f(x).
\end{equation}
Therefore, the two-point time Mellin-like transform can be written as
\begin{equation}
    {\cal I}^{\lambda_1\lambda_2}_{\eta_1\eta_2}[f] = \int_{-\infty}^\infty dx K^{\lambda_1\lambda_2}_{
    \eta_1\eta_2}(x)f(x),
\end{equation}
where the kernel is given by
\begin{equation}
    K^{\lambda_1\lambda_2}_{
    \eta_1\eta_2}(x)\equiv |x|(x+i\eta_1\epsilon)^{-\lambda_1}(x+i\eta_2\epsilon)^{-\lambda_2}I^{\operatorname{sgn}(x)}_{\eta_1\eta_2}(\lambda_1,\lambda_2). 
\end{equation}

The dependence of the kernel on $\sigma$ suggests breaking up ${\cal I}^{\lambda_1\lambda_2}_{\eta_1\eta_2}[f]$ into positive and negative $x$ integrals
\begin{equation}
    \begin{split}\label{eq:2pt-integral-1}
        {\cal I}^{\lambda_1\lambda_2}_{\eta_1\eta_2}[f] &= \int_{-\infty}^0 dx |x|(x+i\eta_1\epsilon)^{-\lambda_1}(x+i\eta_2\epsilon)^{-\lambda_2}I^{-}_{\eta_1\eta_2}(\lambda_1,\lambda_2)f(x)\\
        &+\int_{0}^\infty dx |x|(x+i\eta_1\epsilon)^{-\lambda_1}(x+i\eta_2\epsilon)^{-\lambda_2}I^{+}_{\eta_1\eta_2}(\lambda_1,\lambda_2)f(x)\\
        &\equiv {\cal I}^{\lambda_1\lambda_2}_{\eta_1\eta_2}[f_-] + {\cal I}^{\lambda_1\lambda_2}_{\eta_1\eta_2}[f_+],
    \end{split}
\end{equation}
where we have defined $f_\pm(x) = f(x)\theta(\pm x)$. In summary, the calculation of the two-point time Mellin-like transform of a function $f(u_1-u_2)$ reduces to evaluating ${\cal I}^{\lambda_1\lambda_2}_{\eta_1\eta_2}[f_\pm]$ and taking the sum. 

We can now evaluate $I^{\pm}_{\eta_1\eta_2}(\lambda_1, \lambda_2)$ by splitting it as
\begin{equation}
    \begin{split}
        I_{\eta_1\eta_2}^{\sigma}(\lambda_1,\lambda_2) &= \int_{-\infty}^{-1}dy \left(1+y+i\sigma\eta_1\epsilon\right)^{-\lambda_1} \left(y+i\sigma\eta_2\epsilon\right)^{-\lambda_2}\\
        & +\int_{-1}^{0}dy \left(1+y+i\sigma\eta_1\epsilon\right)^{-\lambda_1} \left(y+i\sigma\eta_2\epsilon\right)^{-\lambda_2}\\
        &+\int_{0}^{1}dy \left(1+y-i\sigma\eta_1\epsilon\right)^{-\lambda_1} \left(y+i\sigma\eta_2\epsilon\right)^{-\lambda_2}\\
        &+\int_{1}^{\infty}dy \left(1+y-i\sigma\eta_1\epsilon\right)^{-\lambda_1} \left(y-i\sigma\eta_2\epsilon\right)^{-\lambda_2} \equiv I_1+I_2+I_3+I_4.
    \end{split}
\end{equation}
Each integral can be related to a beta function integral upon a change of variables as follows. For $I_1$ we set $u=-\frac{1}{y}$, then
\begin{equation}
    \begin{split}
        I_1 &= \int_{-\infty}^{-1}dy (1+y+i\sigma\eta_1\epsilon)^{-\lambda_1}(y+i\sigma\eta_2\epsilon)^{-\lambda_2} =\int_{0}^1 \frac{du}{u^2}\left(\frac{u-1}{u}+i\sigma\eta_1\epsilon\right)^{-\lambda_1}\left(-\frac{1}{u}+i\sigma\eta_2\epsilon\right)^{-\lambda_2}\\
        &=(-1+i\sigma\eta_1\epsilon)^{-\lambda_1}(-1+i\sigma\eta_2\epsilon)^{-\lambda_2}\int_0^1\dfrac{du}{u^{2-\lambda_1-\lambda_2}}(1-u)^{-\lambda_1} \\
        &=e^{-i\pi\sigma\eta_1\lambda_1-i\pi\sigma\eta_2\lambda_2}B(\lambda_1+\lambda_2-1,1-\lambda_1).
    \end{split}
\end{equation}
For $I_2$  we set $u=-y$,
\begin{equation}
    \begin{split}
        I_2&=\int_{-1}^0 dy(1+y+i\sigma\eta_1\epsilon)^{-\lambda_1}(y+i\sigma\eta_2\epsilon)^{-\lambda_2} =\int_{0}^1 du (1-u+i\sigma\eta_1\epsilon)^{-\lambda_1}(-u+i\sigma\eta_2\epsilon)^{-\lambda_2}\\
        &=(1+i\sigma\eta_1\epsilon)^{-\lambda_1}(-1+i\sigma\eta_2\epsilon)^{-\lambda_2}\int_0^1 du (1-u)^{-\lambda_1}u^{-\lambda_2}
        =e^{-i\pi\sigma\eta_2\lambda_2}B(1-\lambda_1,1-\lambda_2).
    \end{split}
\end{equation}
For $I_3$ we set $u = \frac{1}{1+y}$,
\begin{equation}
    \begin{split}
        I_3 &= \int_0^1 dy(1+y-i\sigma\eta_1\epsilon)^{-\lambda_1}(y+i\sigma\eta_2\epsilon)^{-\lambda_2} = \int_{1/2}^{1} \frac{du}{u^2} \left(\frac{1}{u}-i\sigma\eta_1\epsilon\right)^{-\lambda_1}\left(\frac{1-u}{u}+i\sigma\eta_2\epsilon\right)^{-\lambda_2}\\
        &=(1-i\sigma\eta_1\epsilon)^{-\lambda_1}(1+i\sigma\eta_2\epsilon)^{-\lambda_2}\int_{1/2}^{1}du\dfrac{1}{u^{2-\lambda_1-\lambda_2}}(1-u)^{-\lambda_2} =\int_{1/2}^{1} du u^{\lambda_1+\lambda_2-2}(1-u)^{-\lambda_2}.
    \end{split}
\end{equation}
Finally for $I_4$ we make the same change of variables, $u= \frac{1}{1+y}$,
\begin{equation}
    \begin{split}
        I_4 &= \int_1^\infty dy(1+y-i\sigma\eta_1\epsilon)^{-\lambda_1}(y-i\sigma\eta_2\epsilon)^{-\lambda_2} = \int_{0}^{1/2} \frac{du}{u^2} \left(\frac{1}{u}-i\sigma\eta_1\epsilon\right)^{-\lambda_1}\left(\frac{1-u}{u}-i\sigma\eta_2\epsilon\right)^{-\lambda_2}\\
        &=(1-i\sigma\eta_1\epsilon)^{-\lambda_1}(1-i\sigma\eta_2\epsilon)^{-\lambda_2}\int_{0}^{1/2}du\dfrac{1}{u^{2-\lambda_1-\lambda_2}}(1-u)^{-\lambda_2} =\int_{0}^{1/2} du u^{\lambda_1+\lambda_2-2}(1-u)^{-\lambda_2},
    \end{split}
\end{equation}
from which we conclude that the sum $I_3+I_4$ is simply
\begin{equation}
    I_3+I_4 = B(\lambda_1+\lambda_2-1,1-\lambda_2).
\end{equation}
Altogether the integral we started with becomes
\begin{equation}
    \begin{split}   I_{\eta_1,\eta_2}^\sigma(\lambda_1,\lambda_2) &= e^{-i\pi\sigma\eta_1\lambda_1-i\pi\sigma\eta_2\lambda_2}B(\lambda_1+\lambda_2-1,1-\lambda_1)+e^{-i\pi\sigma\eta_2\lambda_2}B(1-\lambda_1,1-\lambda_2)\\
    &+B(\lambda_1+\lambda_2-1,1-\lambda_2),
    \end{split}
\end{equation}
which simplifies to
\begin{equation}\label{eq:2pt-integral-result}
    \begin{split}
I_{\eta_1\eta_2}^\sigma(\lambda_1,\lambda_2) &= 2\pi i\dfrac{ \Gamma(\lambda_1+\lambda_2-1)}{\Gamma(\lambda_1)\Gamma(\lambda_2)}\begin{cases}
            0& \eta_1=\eta_2\\
            -\sigma\eta_1 e^{-i\pi\sigma\eta_2\lambda_2},& \eta_1=-\eta_2.
        \end{cases}
    \end{split}
\end{equation}

As a final step, we evaluate \eqref{eq:2pt-integral-1}. Using the formula for $f$ found in \eqref{eq:two-point-flat}, we have for $\eta_1 = -\eta_2$
\be 
\begin{split}
\mathcal{I}^{\lambda_1\lambda_2}_{\eta_1\eta_2} &= -2\pi i \eta_1  2^{d-1} R^{2\Delta-d+1}\dfrac{\Gamma(2\Delta- d+1)}{\Gamma(\Delta-\frac{d}{2}+1)^2} \dfrac{ \Gamma(\lambda_1+\lambda_2-1)}{\Gamma(\lambda_1)\Gamma(\lambda_2)}  e^{-i \pi \Delta \eta_1\zeta}\delta_{S^{d-1}}(z_1, z_2) \\
&\times \int_{-\infty}^{\infty} dx |x|^d (x+i\eta_1 \epsilon)^{-\lambda_1} (x-i\eta_1\epsilon)^{-\lambda_2} \frac{e^{-\frac{i\pi}{2}\zeta (2\Delta - d+1) \sgn{(x)}}}{|x|^{2\Delta}}{\rm sgn}(x) e^{i\pi {\rm sgn}(x) \eta_1 \lambda_2}\\
&= F_{\Delta}(z_1,z_2)\left[ -e^{\frac{i\pi}{2}\zeta(\lambda_1 + \lambda_2)}e^{i\pi \eta_1 \lambda_2} + e^{-i\pi \eta_1 \lambda_{1}} e^{-\frac{i\pi}{2}\zeta(\lambda_1 + \lambda_2)} \right] \delta(d+1 - \lambda_1 - \lambda_2 - 2 \Delta)\\
&= \delta(d+1 - \lambda_1 - \lambda_2 - 2 \Delta)\begin{cases}
    0, \quad \zeta = -\eta_1,\\
    -F_{\Delta}(z_1,z_2) e^{\frac{i\pi}{2}\eta_1(-\lambda_1 + \lambda_2)} 2i \sin \pi \eta_1 (\lambda_1 + \lambda_2), \quad \zeta = \eta_1,
\end{cases}
\end{split}
\ee
where we defined
\be 
F_{\Delta}(z_1,z_2) = -\eta_1 2^{d-1} R^{2\Delta-d+1}4 \pi^2 i \dfrac{ \Gamma(\lambda_1+\lambda_2-1)}{\Gamma(\lambda_1)\Gamma(\lambda_2)} \dfrac{\Gamma(2\Delta- d+1)}{\Gamma(\Delta-\frac{d}{2}+1)^2} e^{-i\pi \Delta} \delta_{S^{d-1}}(z_1, z_2).
\ee
Using the Euler identity
\be 
\Gamma(x) \Gamma(1 - x) = \frac{\pi}{\sin \pi x},
\ee
and using the constraint imposed on the dimensions by the delta function, we find
\be 
\begin{split}
\mathcal{I}^{\lambda_1\lambda_2}_{\eta_1\eta_2} &= e^{-i\pi \Delta} e^{\frac{i\pi}{2}\eta_1(\lambda_2 - \lambda_1)} \frac{8 \pi^3 2^{d-1} R^{2\Delta - d + 1} }{\Gamma(\lambda_1)\Gamma(\lambda_2) \Gamma(\Delta - \frac{d}{2} + 1)^2} \delta_{S^{d-1}}(z_1,z_2)\delta(d+1 - \lambda_1 - \lambda_2 - 2 \Delta),
\end{split}
\ee
or in terms of the CCFT dimension \cite{de2023celestial}
\be 
\label{eq:ccft-dim}
\Delta_{\rm CCFT} = \Delta+\lambda-1,
\ee
\be 
\begin{split}
\mathcal{I}^{\lambda_1\lambda_2}_{\eta_1\eta_2} &= e^{-i\pi \Delta} e^{\frac{i\pi}{2}\eta_1(\lambda_2 - \lambda_1)}  \frac{8 \pi^3 2^{d-1} R^{2\Delta - d + 1} }{\Gamma(\lambda_1)\Gamma(\lambda_2) \Gamma(\Delta - \frac{d}{2} + 1)^2} \delta_{S^{d-1}}(z_1,z_2)\delta(\Delta_{\rm CCFT}^1+\Delta_{\rm CCFT}^2 -d+1).
\end{split}
\ee

\subsection{Shifting the dimensions and setting $u_i\to 0$}\label{app:dim-reduction-ui->0}

In this appendix we apply \eqref{eq:new-prescription} to the two-point function. The large $R$ expansion of the time-ordered correlator reduces to the Wightman function with $\zeta=\eta_1$. This means that $W$ in \eqref{eq:new-prescription} is
\begin{equation}
\label{eq:time-ordered0}
    \begin{split}
        W_{\Delta_1,\Delta_2}(P_1,P_2) &= \dfrac{2^{d-1}e^{-i\pi \frac{\Delta_1+\Delta_2}{2} \theta(-\eta_1\eta_2)}R^{1+\Delta_1+\Delta_2-d}}{\Gamma(1-\frac{d}{2}+\Delta_1)\Gamma(1-\frac{d}{2}+\Delta_2)}\dfrac{(-\eta_1 i)^{\Delta_1+\Delta_2-d+1}\Gamma(\Delta_1+\Delta_2-d+1)}{(u_{12}-i\eta_1 \epsilon)^{\Delta_1+\Delta_2-d+1}}\\
        &\times \delta_{S^{d-1}}(\Omega_1,\Omega_2) + \cdots,\,
    \end{split}
\end{equation}
which is obtained from \eqref{eq:2pt-expansion-complete} by setting $\eta_1=\zeta$ and recalling that $\Delta = \Delta_1 = \Delta_2$. Following \eqref{eq:new-prescription} we first set $\Delta_i\to \Delta_i^{\rm CCFT}$ and then take the limit $u_i\to 0$. To take the limit we use the identity
\begin{equation}
    \lim_{u_{12}\to 0}\dfrac{(-\eta i)^{\alpha}\Gamma(\alpha)}{(u_{12}-i\eta \epsilon)^{\alpha}}= \lim_{u_{12}\to 0} \int_0^\infty d\omega \omega^{\alpha-1}e^{-i\eta \omega u_{12}} = \int_0^\infty d\omega \omega^{{\alpha-1}} = 2\pi \delta(\alpha),\quad  \alpha\in i\mathbb{R}\,.
\end{equation}
Observe that after shifting $\Delta_i\to \Delta_i^{\rm CCFT}$ in \eqref{eq:time-ordered0},  $\alpha = \Delta_1^{\rm CCFT}+\Delta_2^{\rm CCFT}-d+1$  and we hence obtain the correct two-point delta function $\delta\left(\Delta_1^{\rm CCFT}+\Delta_2^{\rm CCFT}-d+1\right)$. As a result, we see that \eqref{eq:time-ordered0} with $\Delta_i\to \Delta_i^{\rm CCFT}$ and $u_1=u_2=0$ becomes 
\begin{equation}
    \begin{split}
        W_{\Delta_1^{\rm CCFT},\Delta_2^{\rm CCFT}}(P_1,P_2) &= \dfrac{2^{d-1}e^{-i\pi\frac{\Delta_1^{\rm CCFT}+\Delta_2^{\rm CCFT}}{2} \theta(-\eta_1\eta_2)}R^{1+\Delta_1^{\rm CCFT}+\Delta_2^{\rm CCFT}-d}}{\Gamma(1-\frac{d}{2}+\Delta_1^{\rm CCFT})\Gamma(1-\frac{d}{2}+\Delta_2^{\rm CCFT})}\\
        &\times 2\pi \delta\left(\Delta_1^{\rm CCFT}+\Delta_2^{\rm CCFT}-d+1\right)\delta_{S^{d-1}}(\Omega_1,\Omega_2)+ \cdots\,.
    \end{split}
\end{equation}
Further recalling that $\eta_1\eta_2=-1$ for the non-vanishing dimensionally reduced two-point functions, 
\begin{equation}
    \begin{split}
        W_{\Delta_1^{\rm CCFT},\Delta_2^{\rm CCFT}}(P_1,P_2) &= \dfrac{i^{-\Delta_1^{\rm CCFT}}i^{-\Delta_2^{\rm CCFT}}C_{\Delta_1^{\rm CCFT}}C_{\Delta_2^{\rm CCFT}}}{\Gamma(\Delta_1^{\rm CCFT})\Gamma(\Delta_2^{\rm CCFT})}\\
        &\times 2(2\pi)^{d+1} \delta\left(\Delta_1^{\rm CCFT}+\Delta_2^{\rm CCFT}-d+1\right) \delta_{S^{d-1}}(\Omega_1,\Omega_2)+ \cdots
    \end{split}
\end{equation}
from which we observe that accounting for the normalization factor in \eqref{eq:new-prescription} we obtain exactly the expected CCFT$_{d-1}$ two-point function \eqref{eq:two-point-final}.

\section{Three-point function: naive attempt}
\label{sec:three-point-naive}

The goal of this appendix is to illustrate some traps that one may encounter when dealing with distributional correlators. Following the successful recovery of the CCFT$_{d-1}$ two-point function from an appropriate expansion of the Lorentzian CFT$_d$ correlator, one may be tempted to apply these results to the three-point correlators. In particular, the Euclidean CFT$_d$ three-point functions
\eqref{eq:3-pt} are simply a product of the two point structures \eqref{eq:two-point} whose analytic continuations to Lorentzian signature have been computed in section \ref{sec:two-point}. We now show that simply applying our prescription to the product of distributions obtained by naively expanding each of the analytically continued factors in \eqref{eq:3-pt} at large $R$ does \textit{not} lead to the expected result. 

As a first step, we analytically continue \eqref{eq:3-pt} to Lorentizan signature by setting $t_i = \eta_i\frac{\pi}{2} + \frac{u_i}{R}$. We then apply the large-$R$ expansion \eqref{eq:2pt-expansion} to the three factors in \eqref{eq:3-pt} after multiplying each operator by the normalizing factor \eqref{eq:normz} according to \eqref{eq:exp}. The result involves a sum over 8 terms that can be grouped according to the number of delta functions they contain. Furthermore, these terms appear multiplied by different powers of $R$ as one can see from \eqref{eq:2pt-expansion} and \eqref{eq:c12}. The scaling of these terms with $R$ is summarized in Table \ref{tab:R-scaling}, assuming that $C_{123} \propto R^{\sigma}$.

We now see that the term involving three delta functions vanishes in the $R\to\infty$ limit, provided that $\sigma < 3\frac{d-1}{2}$, while the terms with no delta function vanish provided that 
$\sigma < \sum_{i = 1}^3 \Delta_i - 3\frac{d-1}{2}$. Finally, under the assumption that $\sigma < 2\Delta_i -\frac{d-1}{2}$, the term with one delta function vanishes in the $R\to \infty$ limit as well. Terms with two delta functions have a chance to survive when some of the combinations $\Delta_i-\Delta_j-\Delta_k = \frac{d-1}{2} - \sigma$. In fact, as we will see later, taking any of the three possible such combinations of dimensions to satisfy this identity, one can see that the other two combinations will suppressed by powers of $R$ for a choice of $\sigma$ obeying the bounds above, and $\Delta_i$ positive (but not necessarily above the unitarity bound). For example, taking $\Delta_1-\Delta_2-\Delta_3=\frac{d-1}{2} - \sigma$, the other terms respectively scale as $R^{1-d-2\Delta_3 {+2\sigma}}$ and $R^{1-d-2\Delta_2 {+ 2\sigma}}$.

For an appropriate choice of $\sigma$, the surviving term will involve the delta function terms from the expansions of $\frac{1}{(-P_1\cdot P_2)^{\frac{\alpha_{12}}{2}}}$ and $\frac{1}{(-P_1\cdot P_3)^{\frac{\alpha_{13}}{2}}}$ combined with the regular term from $\frac{1}{(-P_1\cdot P_2)^{\frac{\alpha_{23}}{2}}}$ yielding
\begin{equation}
\label{3-pt-Wightman}
    \begin{split}
 \wightmanthree{1}{2}{3} \equiv    \left(\prod_{i=1}^3 \Ni^{-1}\right){\cal W}_{\{\zeta_{ij}\}}(P_1,P_2,P_3) &= {\cal A} \dfrac{e^{-\frac{i\pi}{2}\left(\alpha_{12} {\zeta_{12}\eta_1}\theta(-\eta_1\eta_2) + \alpha_{13}{\zeta_{13}\eta_1} \theta(-\eta_1\eta_3)+\alpha_{23} {\zeta_{23}\eta_2} \theta(-\eta_2\eta_3) \right)}}{\left(u_1-u_2-i\zeta_{12}\epsilon\right)^{\alpha_{12} - d+1}\left(u_1-u_3 -i\zeta_{13}\epsilon\right)^{\alpha_{13}-d+1}}\\
                &\times e^{-\frac{i\pi}{2}\left(\zeta_{12} \alpha_{12} + \zeta_{13} \alpha_{13} - 2d + 2\right)}\frac{\delta_{S^{d-1}}(z_1,z_2)\delta_{S^{d-1}}(z_1,z_3)}{|\Omega_2-\Omega_3|^{\alpha_{23}}}.
    \end{split}
\end{equation}
Here $\mathcal{A}$ combines the contributions from the normalization factors \eqref{eq:normz} and is given by
\be 
\begin{split}
\mathcal{A} &= 2^{\frac{1}{2}\left(\Delta_1 + \Delta_2 + \Delta_3\right)} \pi^{\frac{5d}{2}-4} i^{\Delta_1 + \Delta_2 + \Delta_3}(\eta_1 i)^{\lambda_1} (\eta_2 i)^{\lambda_2} (\eta_3 i)^{\lambda_3}\Gamma\left(\Delta_1 - \frac{d}{2}+1\right)\Gamma\left(\Delta_2 - \frac{d}{2}+1\right)\\
&\times \Gamma\left(\Delta_3 - \frac{d}{2}+1\right) \Gamma(\lambda_1) \Gamma(\lambda_2) \Gamma(\lambda_3) \frac{\Gamma(\frac{\alpha_{12}}{2} - \frac{d-1}{2})\Gamma(\frac{\alpha_{13}}{2} - \frac{d-1}{2})}{\Gamma(\alpha_{12}/2)\Gamma(\alpha_{13}/2)} C_{123}.
\end{split}
\ee
Similar structures with $ 1 \leftrightarrow 2$ and $1 \leftrightarrow 3$ can be obtained by respectively setting $\Delta_2 - \Delta_1 - \Delta_3 = \frac{d-1}{2} {- \sigma}$ , $\Delta_3 - \Delta_1 - \Delta_2 = \frac{d-1}{2}{- \sigma}$. 

Note that such three point functions will always be dominant in unitary CFT$_d$ provided that $\sigma \leq -\frac{1}{2}$. In holographic CFTs, we find that $\sigma = \frac{5 - d}{2}$ (see footnote \ref{fn:C-scale}),
so for low $d$ it is not guaranteed that only the expected celestial three-point structure will survive at large $R$.
We now assume that the dimensions of the CFT$_d$ operators have been chosen such that only  structures with two delta functions survive in the large-$R$ limit. We will a-posteriori verify that for the choice that allows us to recover the Euclidean CCFT$_{d-1}$ amplitudes, all other structures will be indeed suppressed. 

Motivated by the results of section \ref{sec:two-point}, we consider the time-ordered correlators 
\be 
\label{eq:T}
\mathcal{T}_{123} =   \wightmanthree{1}{2}{3} \theta(\tau_1-\tau_2)\theta(\tau_2-\tau_3) + {\rm permutations.}
\ee
 Here $\mathcal{W}_{ijk}$ is the analytic continuation of \eqref{eq:3-pt} to Lorentzian signature. 
Moreover, given $\tau_i = \frac{\eta_i\pi}{2}+\frac{u_i}{R}$ and denoting $u_{ij}=u_i-u_j$ we have
\begin{equation}
\label{eq:theta-fc}
    \theta(\tau_i-\tau_j) = \theta\left((\eta_i-\eta_j)\frac{\pi}{2}+\frac{u_{ij}}{R}\right).
\end{equation}
Using \eqref{eq:theta-exp} we find the large-$R$ expansions of \eqref{eq:T} to be of the form
\be 
\label{eq:time-ordered}
\begin{split}
 {\cal T}_{123}^{+++} &= \wightmanthree{1}{2}{3} \theta(u_{12})\theta(u_{23})+ \wightmanthree{1}{3}{2}\theta(u_{13})\theta(u_{32})+ \wightmanthree{2}{1}{3}\theta(u_{21})\theta(u_{13})\\
 &+ \wightmanthree{2}{3}{1}\theta(u_{23})\theta(u_{31})+ \wightmanthree{3}{1}{2}\theta(u_{31})\theta(u_{12})+ \wightmanthree{3}{2}{1}\theta(u_{32})\theta(u_{21}),\\
\mathcal{T}_{123}^{++-} &= \wightmanthree{1}{2}{3}\theta(u_{12})+ \wightmanthree{2}{1}{3}\theta(u_{21}),\\
{\cal T}_{123}^{-++} &= \wightmanthree{2}{3}{1}\theta(u_{23})+ \wightmanthree{3}{2}{1}\theta(u_{32}),
\end{split}
\ee
with the other in/out configurations related by permutations and crossing symmetry, and where the component of the Wightman function involving two delta functions was given in \eqref{3-pt-Wightman}. The superscripts indicate whether the corresponding operator has been inserted on a future ($\eta_i = 1$) or past ($\eta_i = -1$) sphere. 

We next integrate the external operators against $\int_{-\infty}^{\infty} du_i (u_i + i\eta_i \epsilon)^{-\lambda_i}$. The resulting correlators with the constant $\mathcal{A}$ stripped off are then linear combinations of 
\begin{equation}
\label{eq:3-pt-int}
\begin{split}
    {\cal I}^{\pm}_{3pt} &= \left(\prod_{i=1}^3 \int_{-\infty}^{\infty} du_i \right)\chi_{\pm}(u_1,u_2,u_3)\dfrac{|u_1|^{-\lambda_1}|u_2|^{-\lambda_2}|u_3|^{-\lambda_3}}{|u_1-u_2|^{\alpha_{12}-d+1}|u_1-u_3|^{\alpha_{13}-d+1}},
    \end{split}
\end{equation}
where for mixed in/out correlators such as $\mathcal{T}^{++-}$, $\chi \in U(1)$ takes the form
\be 
\label{eq:phase-main}
\begin{split}
\chi_{\pm}(u_1,u_2,u_3) &= e^{-\frac{i\pi}{2}\left({\zeta_{12}}(\alpha_{12} - d+1)\sgn{(u_{12})} +{\zeta_{13}}(\alpha_{13} - d+1)\sgn{(u_{13})} \right)} \\
&\times e^{-\frac{i\pi}{2}\left(\alpha_{12}{\eta_1\zeta_{12}} \theta(-\eta_1\eta_2) + \alpha_{13} {\eta_1\zeta_{13}} \theta(-\eta_1\eta_3)+\alpha_{23} {\eta_2\zeta_{23}} \theta(-\eta_2\eta_3) \right)}\\
&\times e^{-i\pi(\eta_1\lambda_1\theta(-u_1) + \eta_2\lambda_2\theta(-u_{2}) + \eta_3\lambda_3\theta(-u_3))}\theta(\pm u_{ij}).
\end{split}
\ee
For the all in/out correlators, the step function is replaced by a product of step functions according to \eqref{eq:time-ordered}. 

To evaluate \eqref{eq:3-pt-int}, we make the change of variables $u_1 = x, u_2 = xy, u_3 = xz$. The integrand then factorizes except in the cases associated with $\mathcal{T}^{\pm\pm\pm}_{123}$ and $\mathcal{T}^{\pm \mp \mp}_{123}$ for which the presence of $\theta(\pm x(y - z))$ naively spoils factorization. Fortunately, in these cases we find that the coefficients of the contributing theta functions are all equal. As a result, the theta functions add up to $1$ and all integrals decouple for $y, z$ in the intervals 
\begin{equation}
\label{eq:int-main}
    \mathscr{I}_1 = (-\infty,0),\quad \mathscr{I}_2 = (0,1),\quad \mathscr{I}_3=(1,\infty).
\end{equation}
The final result can be put into the form
\begin{equation}
\begin{split}
    {\cal I}_{3pt}^{\eta_1\eta_2\eta_3} &= 2\pi N_{3pt}\delta(-\alpha_{12}-\alpha_{13}-\lambda_1-\lambda_2-\lambda_3+2d+1),
    \end{split}
\end{equation}
where 
\begin{equation}
\begin{split}
\label{eq:N3pt}
   & N_{3pt}^\pm = \sum_{i,j}\chi_\pm^{ij}\int_{\mathscr{I}_i}dy\dfrac{|y|^{-\lambda_2}}{|1-y|^{\alpha_{12}-d+1}}\int_{\mathscr{I}_j}dz\dfrac{|z|^{-\lambda_3}}{|1-z|^{\alpha_{13}-d+1}},\quad N_{3pt} \equiv N_{3pt}^{+} + N_{3pt}^-
    \end{split}
\end{equation}
and $\chi^{ij}_{\pm}$ is the restriction of \eqref{eq:phase-main} to the interval $\mathscr{I}_i \times \mathscr{I}_j$. Further including the normalization of the operators \eqref{eq:normz}, we find
\begin{equation}
\label{eq:int-CFT-3-pt-main}
    \prod_{i=1}^3 \Ni^{-1}\int_{-\infty}^{\infty}du_i (u_i+i\eta_i\epsilon)^{-\lambda_i} {\cal T}^{\eta_1\eta_2\eta_3}_{123} = {\cal C}_{\Delta_1\Delta_2\Delta_3}^{\eta_1\eta_2\eta_3}(\lambda_1,\lambda_2,\lambda_3) \dfrac{\delta_{S^{d-1}}(z_1,z_2)\delta_{S^{d-1}}(z_1,z_3)}{|\Omega_2-\Omega_3|^{\alpha_{23}}}, 
\end{equation}
where the three-point coefficient is given by
\begin{equation}
    {\cal C}_{\Delta_1\Delta_2\Delta_3}^{\eta_1\eta_2\eta_3}(\lambda_1,\lambda_2,\lambda_3) =  {\cal A} \mathcal{I}_{3pt}^{\eta_1\eta_2\eta_3}
\end{equation}
and $\mathcal{A}$ includes the normalizing factors. The dependence of $\mathcal{A}$ and $\widetilde{\mathcal{A}}$ on the dimensions is left implicit. It will be convenient to further decompose the coefficient into $\lambda_i$ dependent and $\lambda_i$ independent terms, namely
\begin{equation}
\label{eq:3-pt-cel-main}
    \begin{split}
     2\pi{\cal A}N_{3pt} &\equiv \widetilde{{\cal A}} e^{\frac{i\pi}{2}(\eta_1\lambda_1+\eta_2\lambda_2+\eta_3\lambda_3)} \Gamma(\lambda_1)\Gamma(\lambda_2+\alpha_{12}-d)\Gamma(\alpha_{13}+\lambda_3-d)\hat{N}_{3pt}^{\eta_1\eta_2\eta_3},
    \end{split}
\end{equation}
where
\begin{equation}
\label{eq:norm-net}
    \begin{split}
    \widetilde{{\cal A}} &=   \dfrac{2^{1+\frac{\Delta_1+\Delta_2+\Delta_3}{2}}\pi^{\frac{5d}{2}-3}i^{\Delta_1+\Delta_2+\Delta_3}\Gamma(\frac{\alpha_{12}}{2}-\frac{d-1}{2})\Gamma(\frac{\alpha_{13}}{2}-\frac{d-1}{2})}{\Gamma(\frac{\alpha_{12}}{2})\Gamma(\frac{\alpha_{13}}{2})\Gamma(\alpha_{12}-d+1)\Gamma(\alpha_{13}-d+1)}\prod_{i=1}^3 \Gamma\left(\Delta_i-\frac{d}{2}+1\right) C_{123}.
    \end{split}
\end{equation}
Note that for $d = 3$, $\widetilde{\mathcal{A}}$ simplifies to
\be 
\label{eq:A}
\widetilde{{\cal A}} =  2^{3+\frac{\Delta_1+\Delta_2+\Delta_3}{2}}\pi^{\frac{9}{2}}i^{\Delta_1+\Delta_2+\Delta_3} \frac{\Gamma(\Delta_1 - \frac{1}{2})\Gamma(\Delta_2 - \frac{1}{2})\Gamma(\Delta_3 - \frac{1}{2})}{\Gamma(\alpha_{12} - 1)\Gamma(\alpha_{13} - 1)} C_{123}.
\ee

We would now like to compare this result to the celestial amplitudes in \eqref{eq:ccft-res-d}. 
 We further note that using the reflection formula we can bring any of the three gamma functions in  $2\pi{\cal A}N_{3pt}$ to the denominator. On the support of the delta function, we can then recognize an Euler beta function and we obtain three equivalent expressions for \eqref{eq:3-pt-cel-main}. The first is 
\begin{equation}\label{eq:beta-function-1}
\begin{split}
     2\pi{\cal A}N_{3pt} &\simeq \widetilde{{\cal A}} e^{\frac{i\pi}{2}(\eta_1\lambda_1+\eta_2\lambda_2+\eta_3\lambda_3)} \dfrac{\pi B(\lambda_2+\alpha_{12}-d,\lambda_3+\alpha_{13}-d)}{\sin \pi \lambda_1}\hat{N}_{3pt}^{\eta_1\eta_2\eta_3},
    \end{split}
\end{equation}
the second is
\begin{equation}\label{eq:beta-function-2}
\begin{split}
     2\pi{\cal A}N_{3pt} &\simeq \widetilde{{\cal A}} e^{\frac{i\pi}{2}(\eta_1\lambda_1+\eta_2\lambda_2+\eta_3\lambda_3)} \dfrac{\pi B(\lambda_1,\alpha_{13}+\lambda_3-d)}{\sin \pi (\alpha_{12}+\lambda_2-d)}\hat{N}_{3pt}^{\eta_1\eta_2\eta_3},
    \end{split}
\end{equation}
while the third is
\begin{equation}\label{eq:beta-function-3}
\begin{split}
     2\pi{\cal A}N_{3pt} &\simeq \widetilde{{\cal A}} e^{\frac{i\pi}{2}(\eta_1\lambda_1+\eta_2\lambda_2+\eta_3\lambda_3)} \dfrac{\pi B(\lambda_1,\alpha_{12}+\lambda_2-d)}{\sin \pi (\alpha_{13}+\lambda_3-d)}\hat{N}_{3pt}^{\eta_1\eta_2\eta_3},
    \end{split}
\end{equation}
where by $\simeq$ we mean equality modulo the delta function constraint. In terms of the CCFT dimensions
\be 
\Delta_{\rm CCFT} = \lambda + \Delta - 1,
\ee
the three Euler beta functions become
\be 
\begin{split}
B(\lambda_2+\alpha_{12}- d,\lambda_3+\alpha_{13}-d) &= B\left(\Delta_{\rm CCFT}^2 - \Delta_2 +1 -d + \alpha_{12},\Delta_{\rm CCFT}^3 - \Delta_3  +1-d + \alpha_{13}\right),\\
    B(\lambda_1,\alpha_{13} + \lambda_3 - d) &= B(\Delta_{\rm CCFT}^1 - \Delta_1 + 1, \Delta_{\rm CCFT}^3 - \Delta_3 +1-d + \alpha_{13}),\\
     B(\lambda_1,\alpha_{12}+\lambda_2-d) &= B(\Delta_{\rm CCFT}^1 - \Delta_1 + 1, \Delta_{\rm CCFT}^2 - \Delta_2 +1-d + \alpha_{12}).
\end{split}
\ee
We now notice that setting $\Delta_1 = d-1, \Delta_2 = \Delta_3 = 1$, or equivalently $\alpha_{12} = \alpha_{13} = d-1$, $\alpha_{23} = 3-d$, the arguments of the beta functions reduce to those of the scalar celestial amplitudes in \eqref{eq:ccft-res-d}. Moreover, we see from \eqref{eq:int-CFT-3-pt-main} that the angular dependence also matches for $\alpha_{23} = 3-d$. It remains to determine and compare the normalization which we do in appendix \ref{app:3-pt-normalization}.

 At this point, we encounter a puzzle. We argued from general grounds in Section \ref{sec:preliminaries} that our prescription should yield celestial amplitudes that only depend on the combination of dimensions $\Delta_i + \lambda_i - 1 \equiv \Delta_i^{\rm CCFT}$. On the other hand, the approach outlined in this appendix yields correlators with further isolated dependence on $\Delta_i$ which only disappears for specific choices of $\Delta_i$. That certain constraints on $\Delta_i$ are needed can already be seen by analyzing the terms in Table \ref{tab:R-scaling} which would all otherwise scale non-trivially with $R$. Furthermore, note that we needed to set some of the dimensions $\Delta_i = 1$ which clearly violate the CFT$_d$ unitarity bounds.\footnote{Despite this, we can check that all but the required flat space CCFT$_{d-1}$ three-point structures are still \textit{marginally} suppressed. In particular, we need to shift $\Delta_1 = d-1 + \epsilon_1, \quad \Delta_2 = \Delta_3 = 1 + \epsilon_2,$ where $\epsilon_1$ and $\epsilon_2$ must be necessarily positive in order for all but the correct conformal three-point structure on $S^{d-1}$ to contribute to \eqref{eq:3-pt} in the large-$R$ limit. It is necessary to take $R \rightarrow \infty$ \textit{before} taking $\epsilon_i \rightarrow 0$ which ensures that only a three-point structure involving two delta functions contributes (see Table \ref{tab:R-scaling}). }

We find that for this choice of external dimensions, $\chi_{\pm}$ simplify drastically {(see appendix \ref{app:3-pt-normalization})} to yield
\be 
\label{eq:norm-app}
\begin{split}
\hat{N}_{3pt}^{+++} &= 0,\\
\hat{N}_{3pt}^{-++} &= e^{i\pi(d+1)} 2i \pi^2 e^{i\pi \lambda_1} \sin \pi \lambda_1,\\
\hat{N}_{3pt}^{++-} &= 2i \pi^2 e^{i\pi \lambda_3} \sin \pi \lambda_3.
\end{split}
\ee

Including the normalization \eqref{eq:norm-net}, we find the non-vanishing 3-point $S^{d-1}$ correlators 
\be
\begin{split}\label{eq:3pt-coeff-results}
    C_{\Delta_1\Delta_2\Delta_3}^{-++}(\lambda_1,\lambda_2,\lambda_3) &= \mathscr{N} B(\Delta_{\rm CCFT}^2 - 1,\Delta_{\rm CCFT}^3-1)\delta\left(\sum_{i = 1}^3 \Delta_{\rm CCFT}^i -d-1\right),\\
C_{\Delta_1\Delta_2\Delta_3}^{++-}(\lambda_1,\lambda_2,\lambda_3) &= (-1)^d \mathscr{N}B(\Delta_{\rm CCFT}^1-d+2,\Delta_{\rm CCFT}^2-1)\delta\left(\sum_{i = 1}^3 \Delta_{\rm CCFT}^i -d-1\right),\\
\mathscr{N}&\equiv \dfrac{(-i)^{d+1}2^{\frac{9+d}{2}}\pi^{\frac{5d}{2}}\Gamma(2-\frac{d}{2})^2\Gamma(\frac{d}{2})C_{123}}{\Gamma(\frac{d-1}{2})^2},
\end{split}\ee
with $\Delta_{\rm CCFT}^i$ defined in \eqref{eq:dccft}. One can explicitly check that the three-point amplitude derived from the flat space limit obeys the constraints of Poincaré invariance discussed in \cite{Law:2019glh}. In addition to the angular dependence required by momentum conservation, the beta function and the Dirac delta function in dimensions ensure the three-point coefficient obeys the recursion relation implied by translation symmetry.

All other configurations are obtained by permutations and crossing. The factor of $(-1)^d$ 
appearing only in the $\eta_2 = \eta_3$ configurations may be understood by noting that the correlators in \eqref{eq:int-CFT-3-pt-main} always contain the transverse delta functions $\delta_{S^{d-1}}(z_1,z_2) \delta_{S^{d-1}}(z_1,z_3)$ whose arguments may involve points on opposite or the same sphere depending on which in/out configuration is considered. On the other hand, the transverse delta functions in \eqref{eq:ccft-res-d} always involve points on antipodally related spheres. For example, when replacing $\delta_{S^{d-1}}(z_1,z_2)$ multiplying $C_{\Delta_1\Delta_2\Delta_3}^{++-}$ by $\delta_{S^{d-1}}(z_2,z_3)$, one acquires a factor of $(-1)^d$ since the antipodal map reverses the orientation of $S^{d-1}$ for odd $d$. We conclude that $\mathscr{N}$ should give the CCFT result provided that $C_{123} = C_{123}^{AdS}$ is the OPE coefficent corresponding to a 3-point contact diagram in AdS$_{d+1}$.

 This OPE coefficient has been computed using tree level AdS Witten diagrams \cite{freedman1999correlation} and, with the normalization conventions we use here it takes the form\footnote{ Our normalization of the bulk-to-boundary propagator \eqref{eq:norm} differs from that in \cite{freedman1999correlation} by $2^{\frac{\Delta_1 + \Delta_2 + \Delta_3}{2}- 3}$. Furthermore, \cite{freedman1999correlation} consider an Euclidean action in AdS in which case the three-point vertex differs by $-i$ from the Lorentzian vertex. However, we have also dropped the factor of $i$ accompanying the three-point scalar interaction in \eqref{eq:flat-three-point}, so we can directly compare \eqref{eq:3pt-coeff-results} to \eqref{eq:ccft-res-d}.} 
    \begin{equation}\label{eq:ads-ope-coeff-unit-radius}
    C_{123}^{AdS} = g{2^{\frac{\Delta_1+\Delta_2+\Delta_3}{2}-{3}}} \frac{\Gamma\left(\frac{\alpha_{12}}{2} \right)\Gamma\left(\frac{\alpha_{13}}{2}\right)\Gamma\left(\frac{\alpha_{23}}{2}\right)\Gamma\left(\frac{\alpha_{12} + \alpha_{13} + \alpha_{23} - d }{2} \right)}{2 \pi^d \Gamma\left(\Delta_1 - \frac{d}{2} + 1\right)\Gamma\left(\Delta_2 - \frac{d}{2} + 1\right)\Gamma\left(\Delta_3 - \frac{d}{2} + 1 \right)}.
\end{equation}
In this case and for the values of $\Delta_i$ needed to reproduce the expected beta functions in flat space $(\Delta_1 = d-1, \Delta_2 = \Delta_3 = 1)$, we find
\begin{equation}
    \begin{split}
    \mathscr{N} &=  {-}g2^{d+1}\pi^{\frac{3(d+1)}{2}}(-i)^{d  + 1}\dfrac{\sec\frac{\pi d}{2}}{\Gamma(\frac{d-1}{2})}.
    \end{split}
\end{equation}
In terms of the CCFT coefficient we have
\begin{equation}
\label{eq:puzzle}
    \begin{split}
   \mathscr{N} &=  -2^{-2}(-i)^{d+1}\sec\frac{\pi d}{2}\times \mathscr{N}_{\rm CCFT}.
    \end{split}
\end{equation}
Comparing to \eqref{eq:ccft-res-d}, this agrees with the CCFT$_{d-1}$ result up to a factor $-\frac{1}{4}(-i)^{d +1}\sec\frac{\pi d}{2}$ that diverges in odd dimensions. 
Hence, the lesson we draw from this computation is that the distributional component of a product of functions may not coincide with the product of distributions.

In Section \ref{sec:three-point} we show how to obtain the expected result by carefully treating the large-$R$ limit of the whole three-point Lorentzian correlator \eqref{eq:euclidean-three-point}. No restriction on the CFT$_d$ operator dimensions, other than that they satisfy unitarity bounds is necessary in this case.

\subsection{Normalization}
\label{app:3-pt-normalization}

In this section we give the details leading to \eqref{eq:norm-app}.
We can compute ${\cal I}_{3pt} = {\cal I}_{3pt}^++{\cal I}_{3pt}^-$, in which ${\cal I}_{3pt}^\pm$ have been defined in \eqref{eq:3-pt-int}, by
\begin{equation}
\label{eq:Ithree-pt}
    {\cal I}_{3pt} = 2\pi(N_{3pt}^++N_{3pt}^-)\delta(-\alpha_{12}-\alpha_{13}-\lambda_1-\lambda_2-\lambda_3+2d+1)\,,
\end{equation}
and $N_{3pt}^\pm$ is given by
\begin{equation}
\label{eq:norm-3-pt}
    N_{3pt}^\pm = \sum_{i,j}\chi_\pm^{ij} \mathcal{I}_i(\lambda_2,\alpha_{12}-d+1)\mathcal{I}_j(\lambda_3,\alpha_{13}-d+1).
\end{equation}
Here we defined the three basic integrals (see \eqref{eq:N3pt})
\begin{equation}
    \begin{split}
        {\cal I}_1(a,b) &= \dfrac{\Gamma(a+b-1)}{\Gamma(a)\Gamma(b)}\dfrac{\pi}{\sin(a\pi)},\\
        {\cal I}_2(a,b) &= -\dfrac{\Gamma(a+b-1)}{\Gamma(a)\Gamma(b)}\left(\pi\cot(a\pi)+\pi\cot(b\pi)\right),\\
        {\cal I}_3(a,b) &= \dfrac{\Gamma(a+b-1)}{\Gamma(a)\Gamma(b)}\dfrac{\pi}{\sin(b\pi)}.
    \end{split}
\end{equation}
Furthermore $\chi_\pm^{ij} = e^{i\Phi_{\eta_1,\eta_2,\eta_3}(x,y,z)}$ with $\operatorname{sgn}(x)=\pm 1$ and $(y,z)\in \mathscr{I}_i\times \mathscr{I}_j$. In the following, we analyze the different in-out configurations.
Specifically, we find 
\begin{equation}
\label{eq:phase}
\begin{split}
    \chi(u_i) &\equiv e^{-\frac{i\pi}{2}\left(\alpha_{12} \theta(-\eta_1\eta_2)+\alpha_{13} \theta(-\eta_1\eta_3)+\alpha_{23} \theta(-\eta_2\eta_3) \right)} e^{-\frac{i\pi}{2}\left(\zeta_{12}(\alpha_{12} - d+1) \sgn{(x(1-y))} +{\zeta_{13}}(\alpha_{13} - d+1) \sgn{(x(1 - z))}\right)} \\
&\times e^{-i\pi\left(\eta_1\lambda_1\theta(-x) + \eta_2\lambda_2\theta(-xy) + \eta_3\lambda_3\theta(-xz) \right)}.
     \end{split}
\end{equation}

We are interested in the case of $\alpha_{12} = \alpha_{13} = 2, \alpha_{23} = 0$ and $d = 3$. (In arbitrary higher dimensions we have $\alpha_{12} = \alpha_{13} = d - 1$ and $\alpha_{23} = 3-d$ for scalars.)  In this case, the phases $\chi_{\pm}^{ij}$ further simplify as the dependence on $\sgn{(x(1-y))}$ and $\sgn{(x(1-z))}$ drops out. 
However, the coefficients in \eqref{eq:norm-3-pt} are individually divergent. It is hence important to evaluate the linear combination in \eqref{eq:norm-3-pt} before setting $d = 3$ and $\alpha_{12} = \alpha_{13} = 2$, $\alpha_{23} = 0$. We note that the phase in \eqref{eq:phase} could in principle depend on a regulator that determines the rate at which $\alpha_{ij}$ approach their respective values. We found the result to be highly sensitive to this regulator (as one would expect given that the coefficients in \eqref{eq:norm-3-pt} diverge). However, we also found naive choices for this regulator (eg. $\alpha_{ij}$ $\rightarrow \alpha_{ij} + \varepsilon$) to spoil the symmetry of the results under operator exchanges. Such corrections may appear naturally in the CFT$_3$ due to quantum effects. At large $N$, they would be suppressed by powers of $N$ or equivalently $R$.  Interestingly, the choice of CFT$_{d}$ operators that gives rise to the scalar amplitudes in CCFT$_{d-1}$ are $\Delta_1 = d - 1$ and $\Delta_2 = \Delta_3 = {1}$. We henceforth assume the phase to be independent on such corrections and leave a better understanding of related effects to future work. 

For the $(\eta_1,\eta_2,\eta_3)=(\pm 1,\pm 1, \pm 1)$ configuration we find that, on the support of the $\theta(u_{ij})$ functions in \eqref{eq:time-ordered}, the phases of all Wightman functions are equal to
\begin{equation}
    \Phi_{\pm 1 \pm 1\pm 1}(x,y,z) = \mp \pi (\lambda_1 \theta(-x)+\lambda_2\theta(-xy) + \lambda_3\theta(-xz))\,.
\end{equation}
This allows $e^{i\Phi_{\pm\pm\pm}(x,y,z)}$ to be factored out, leaving just the sum of $\theta$ functions, that can be shown to sum up to one. Calculating $N_{3pt}^\pm$, we find ${\cal I}_{3pt} = 0$ for both $\eta = \pm 1$ on the support of the delta function in \eqref{eq:Ithree-pt}. This is true for arbitrary $\alpha_{ij}$ on the support of the delta function in \eqref{eq:Ithree-pt}. In the following subsections we assume that $\alpha_{12} = \alpha_{13} = d - 1$ and $\alpha_{23} = 3 - d$.

For the $(\eta_1, \eta_2,\eta_3) = (1,1,-1)$ configuration, the phase function is $\Phi_{+1,+1,-1}(x,y,z)$ given by
\begin{equation}
    \begin{split}
        \Phi_{+1,+1,-1}(x,y,z) &= -\pi(\lambda_1\theta(-x)+\lambda_2\theta(-xy) - \lambda_3\theta(-xz)) - \pi 
    \end{split}
\end{equation}
and we find 
\begin{equation}
\label{eq:11m10}
    \begin{split}
   \hat{N}_{3pt}^{++-} &= 
     \dfrac{4\pi^2 e^{i\pi(d - \lambda_1 - \lambda_2)}(e^{i\pi(\alpha_{12} + \lambda_1 + \lambda_2)} + e^{i\pi(\alpha_{13} + \lambda_3)})}{(e^{i\pi d}-e^{i\pi\alpha_{12}}) (e^{i \pi d} - e^{i\pi \alpha_{13}})}\,\\
     &\simeq  2i \pi^2 e^{i\pi \lambda_3} \sin \pi \lambda_3.
    \end{split}
\end{equation}

 The $(\eta_1, \eta_2,\eta_3) = (1,-1,1)$ configuration can be easily obtained from \eqref{eq:11m10} by $2 \leftrightarrow 3$ yielding 
\begin{equation}\label{eq:11m10-corrected}
    \begin{split}
    \hat{N}_{3pt}^{+-+} 
    &=   \dfrac{4\pi^2 e^{i\pi(d - \lambda_1 - \lambda_3)}(e^{i\pi(\alpha_{13} + \lambda_1 + \lambda_3)} + e^{i\pi(\alpha_{12} + \lambda_2)})}{(e^{i\pi d}-e^{i\pi\alpha_{13}}) (e^{i \pi d} - e^{i\pi \alpha_{12}})}\,\\
     &\simeq  2i \pi^2 e^{i\pi \lambda_2} \sin \pi \lambda_2\,.
    \end{split}
\end{equation}

For $(\eta_1,\eta_2,\eta_3)=(-1,1,1)$ the phase function is $\Phi_{-1,+1,+1}(x,y,z)$ given by
\begin{equation}
    \begin{split}
        \Phi_{-1,+1,+1}(x,y,z) &= \pi(\lambda_1\theta(-x)-\lambda_2\theta(-xy)-\lambda_3\theta(-xz)) - \pi(d-1) \, ,
    \end{split}
\end{equation}
and we find
\begin{equation}
    \begin{split}
    \hat{N}_{3pt}^{-++} &= \frac{4\pi^2 e^{-i\pi (d + \lambda_2+ \lambda_3)}(e^{i\pi(2d+\lambda_1)}+e^{i\pi(\alpha_{12}+\alpha_{13}+\lambda_2+\lambda_3)})}{(e^{i\pi d}-e^{i\pi\alpha_{12}})(e^{i\pi d}-e^{i\pi\alpha_{13}})}\\
     &\simeq  e^{i\pi(d+1)}2i \pi^2 e^{i\pi \lambda_1} \sin \pi \lambda_1.
    \end{split}
\end{equation}

For the $(\eta_1,\eta_2,\eta_3)=(-1,-1,1)$ configuration the phase function is $\Phi_{-1,-1,+1}(x,y,z)$ given by
\begin{equation}
    \begin{split}
        \Phi_{-1,-1,+1}(x,y,z) &= -\pi + \pi(\lambda_1\theta(-x)+\lambda_2\theta(-xy)-\lambda_3\theta(-xz))\,
    \end{split}
\end{equation}
and 
\begin{equation}
\label{eq:mmp0}
    \begin{split}
   \hat{N}_{3pt}^{--+} &= 
     \dfrac{4\pi^2 e^{ i\pi(d -  \lambda_3)}(e^{i\pi(\alpha_{12} + \lambda_1 + \lambda_2)} + e^{i\pi(\alpha_{13} + \lambda_3)})}{(e^{i\pi d}-e^{i\pi\alpha_{12}}) (e^{i \pi d} - e^{i\pi \alpha_{13}})}\,\\
     &\simeq  -2i \pi^2 e^{-i\pi \lambda_3} \sin \pi \lambda_3.
    \end{split}
\end{equation}

As before, the $(\eta_1,\eta_2,\eta_3)=(-1,1,-1)$ configuration is related to \eqref{eq:mmp0} by $2\leftrightarrow 3$ yielding 
\begin{equation}
    \begin{split}
   \hat{N}_{3pt}^{-+-} &= 
     \dfrac{4\pi^2 e^{i\pi (d - \lambda_2)}(e^{i\pi(\alpha_{13} + \lambda_1 + \lambda_3)} + e^{i\pi(\alpha_{12} + \lambda_2)})}{(e^{i\pi d}-e^{i\pi\alpha_{13}}) (e^{i \pi d} - e^{i\pi \alpha_{12}})}\,\\
     &\simeq  -2i \pi^2 e^{-i\pi \lambda_2} \sin \pi \lambda_2.
    \end{split}
\end{equation}

Finally, for the $(\eta_1,\eta_2,\eta_3)=(1,-1,-1)$ configuration the phase function is $\Phi_{1,-1,-1}(x,y,z)$ given by
\begin{equation}
    \begin{split}
        \Phi_{1,-1,-1}(x,y,z) &= -\frac{\pi}{2}(\alpha_{12} + \alpha_{13})-\pi(\lambda_1\theta(-x)-\lambda_2\theta(-xy)-\lambda_3\theta(-xz)),
    \end{split}
\end{equation}
and we find
\begin{equation}
    \begin{split}
    \hat{N}_{3pt}^{+--} &= 
     \dfrac{4 \pi^2 e^{- i\pi d}(1 + e^{i\pi(\alpha_{12} + \alpha_{13} +\lambda_2 + \lambda_3 - \lambda_1)})}{(e^{i\pi d}-e^{i\pi\alpha_{12}}) (e^{i \pi d} - e^{i\pi \alpha_{13}})}\,\\
     &\simeq  -{ e^{i\pi(d+1)}} 2i \pi^2 e^{-i\pi \lambda_1} \sin \pi \lambda_1.
    \end{split}
\end{equation}
We note that the results related by exchanging ${\rm in} \leftrightarrow {\rm out}$ differ only by a phase. We will see below that multilpication by the normalization relating CFT and CCFT operators leads to identical results as expected from crossing symmetry. 

We can now put everything together. 
Using \eqref{eq:beta-function-1} we find that
\begin{equation}
    \begin{split}
        2\pi {\cal A}N_{3pt}^{-++} &\simeq \widetilde{{\cal A}} e^{\frac{i\pi}{2}(-\lambda_1+\lambda_2+\lambda_3)} \dfrac{\pi B(\lambda_2-1,\lambda_3-1)}{\sin \pi \lambda_1}\times 2i\pi^2 {e^{i\pi (d+1)}}e^{i\pi\lambda_1}\sin\pi\lambda_1\\
        &\simeq {(-1)^{d+1}}2\pi^3 \widetilde{{\cal A}} B(\lambda_2-1,\lambda_3-1),
    \end{split}
\end{equation}
and by the same argument
\begin{equation}
    \begin{split}
        2\pi {\cal A}N_{3pt}^{+--} &\simeq {(-1)^{d+1}}2\pi^3\widetilde{{\cal A}} B(\lambda_2-1,\lambda_3-1).
    \end{split}
\end{equation}
Now using \eqref{eq:beta-function-2},
\begin{equation}
    \begin{split}
        2\pi{\cal A}N_{3pt}^{+-+} &\simeq \widetilde{{\cal A}} e^{\frac{i\pi}{2}(\lambda_1-\lambda_2+\lambda_3)} \dfrac{\pi B(\lambda_1,\lambda_3-1)}{\sin \pi (\lambda_2-1)}\times 2i\pi^2 e^{i\pi \lambda_2} \sin \pi \lambda_2\\
    &\simeq -2\pi^3 \widetilde{{\cal A}} B(\lambda_1,\lambda_3-1),
    \end{split}
\end{equation}
and likewise
\begin{equation}
\begin{split}
     2\pi{\cal A}N_{3pt}^{-+-} &\simeq -\widetilde{{\cal A}} e^{\frac{i\pi}{2}(-\lambda_1+\lambda_2-\lambda_3)} \dfrac{\pi B(\lambda_1,\lambda_3-1)}{\sin \pi (\lambda_2-1)}\times 2i\pi^2 e^{-i\pi\lambda_2} \sin \pi \lambda_2\\
     &\simeq -2\pi^3 \widetilde{{\cal A}}B(\lambda_1,\lambda_3-1).
    \end{split}
\end{equation}
Finally using \eqref{eq:beta-function-3} we find
\begin{equation}
\begin{split}
     2\pi{\cal A}N_{3pt}^{++-} &\simeq \widetilde{{\cal A}} e^{\frac{i\pi}{2}(\lambda_1+\lambda_2-\lambda_3)} \dfrac{\pi B(\lambda_1,\lambda_2-1)}{\sin \pi (\lambda_3-1)}\times 2i \pi^2 e^{i\pi \lambda_3} \sin \pi \lambda_3\\
     &\simeq -2\pi^3 \widetilde{{\cal A}} B(\lambda_1,\lambda_2-1),
    \end{split}
\end{equation}
and similarly,
\begin{equation}
\begin{split}
     2\pi{\cal A}N_{3pt}^{--+} &\simeq -\widetilde{{\cal A}} e^{\frac{i\pi}{2}(-\lambda_1-\lambda_2+\lambda_3)} \dfrac{\pi B(\lambda_1,\alpha_{12}+\lambda_2-d)}{\sin \pi (\alpha_{13}+\lambda_3-d)}\times 2i \pi^2 e^{-i\pi \lambda_3} \sin \pi \lambda_3\\
     &\simeq -2\pi^3 \widetilde{{\cal A}} B(\lambda_1,\lambda_2-1).
    \end{split}
\end{equation}

This gives us the complete list of CCFT$_{d-1}$ three-point coefficients derived from Lorentzian CFT$_d$ on the cylinder
\begin{equation}
    \begin{split}
     {\cal C}_{\Delta_1\Delta_2\Delta_3}^{\pm 1, \pm 1 ,\pm 1} &= 0,\\
     {\cal C}_{\Delta_1\Delta_2\Delta_3}^{\mp 1, \pm 1 ,\pm 1} &= (-1)^{d+1} 2\pi^3 \widetilde{{\cal A}}  B(\lambda_2-1,\lambda_3-1)\delta(\lambda_1+\lambda_2+\lambda_3-3),\\
     {\cal C}_{\Delta_1\Delta_2\Delta_3}^{\pm 1, \mp 1 ,\pm 1} &= -2\pi^3\widetilde{{\cal A}} B(\lambda_1,\lambda_3-1)\delta(\lambda_1+\lambda_2+\lambda_3-3),\\
     {\cal C}_{\Delta_1\Delta_2\Delta_3}^{\pm 1, \pm 1 ,\mp 1} &= -2\pi^3\widetilde{{\cal A}} B(\lambda_1, \lambda_2-1)\delta(\lambda_1+\lambda_2+\lambda_3-3).
    \end{split}
\end{equation}
 These can be rewritten in terms of the effective CCFT dimensions \eqref{eq:ccft-dim}, and simplified noting that for $\Delta_1=d-1$ and $\Delta_2=\Delta_3=1$
\be 
\widetilde{\mathcal{A}} = \dfrac{i^{d+1}2^{\frac{7+d}{2}}\pi^{\frac{5d}{2}-3}\Gamma(2-\frac{d}{2})^2\Gamma(\frac{d}{2})}{\Gamma(\frac{d-1}{2})^2}C_{123}.
\ee

We can compare this to the celestial scalar 3-point function using the OPE coefficient obtained from a contact 3-point scalar interaction in AdS \cite{freedman1999correlation}. Following the argument in footnote \ref{fn:C-scale}, 
\begin{equation}
    C_{123}(R) = R^{\frac{5-d}{2}}\hat{C}_{123}.
\end{equation}
Finally by the calculation of \cite{freedman1999correlation} we have
\be 
\begin{split}
\hat{C}_{123} = g \frac{2^{\frac{\Delta_1+\Delta_2+\Delta_3}{2}-3}\Gamma\left(\frac{\alpha_{12}}{2} \right)\Gamma\left(\frac{\alpha_{13}}{2}\right)\Gamma\left(\frac{\alpha_{23}}{2}\right)\Gamma\left(\frac{\alpha_{12} + \alpha_{13} + \alpha_{23} - d }{2} \right)}{2 \pi^d \Gamma\left(\Delta_1 - \frac{d}{2} + 1\right)\Gamma\left(\Delta_2 - \frac{d}{2} + 1\right)\Gamma\left(\Delta_3 - \frac{d}{2} + 1 \right)},
\end{split}
\ee
which for $\alpha_{12}= \alpha_{13} = d - 1$, $\alpha_{23} = 3 - d$ simplifies to
\be 
\hat{C}_{123} = g\dfrac{2^{\frac{1+d}{2}-4}\pi^{\frac{1}{2}-d}\Gamma(\frac{3-d}{2})\Gamma(\frac{d-1}{2})^2}{\Gamma(2-\frac{d}{2})^2\Gamma(\frac{d}{2})}.
\ee

\section{Three-point function: integrating over the strips}
\label{app:3pt-time-mellin}

In this appendix we put \eqref{eq:three-point-int} in a CCFT$_{d-1}$ conformal primary basis by directly integrating it against the $u_i$-dependent kernels. For simplicity, we focus on
\be 
\begin{split}
I &= \prod_{i = 1}^3 \int_{-\infty}^{\infty} du_i \left(u_i + i\eta_i \epsilon \right)^{-\lambda_i} e^{- i\eta_1 \eta_2 \omega_1 \omega_2 u_{12}^2 + {\rm perms.}}\\
& = \prod_{i = 1}^3 \left(\frac{1}{\Gamma(\lambda_i)(\eta_i i)^{\lambda_i}} \int_0^{\infty}
 ds_i s_i^{\lambda_i - 1} \int_{-\infty}^{\infty} du_i e^{i\eta_i s_i u_i} \right) e^{-i\eta_1 \eta_2 \omega_1 \omega_2 u_{12}^2 + {\rm perms.}}\,,
 \end{split}
\ee
 where the dependence on various $i\epsilon$-dependent terms in the exponent is left implicit and will be discussed further below.
The exponent can be written as
\be 
e^{i u^T A u + B^T u},
\ee
where $A$ is a matrix with entries $A_{ij} = -\eta_i \eta_j \omega_i \omega_j$ on the support of the momentum conserving delta function and 
\be 
B = \left( \eta_1 s_1 + i\epsilon_1(\omega_i), \eta_2 s_2 + i\epsilon_2(\omega_i), \eta_3 s_3 + i\epsilon_3(\omega_i)  \right).
\ee
The $i\epsilon$ shifts are inherited from $(u_{ij} \pm i\epsilon)^2$ and depend on $\omega_i$. Depending on the in/out configuration, two of these shifts are positive and one negative or vice-versa. The changes of variables performed below will further complicate the dependence on $\epsilon$. As we will see, the final answer will depend on the choice of branch cuts in the complex $u_i$ space and hence the path chosen for the integrals in the uplift of this space to its universal cover. We will discuss various choices at the end, with the upshot that they will give answers differing by trigonometric functions in the CFT$_d$ operator dimensions.

\begin{figure}
\includegraphics[scale=0.45]{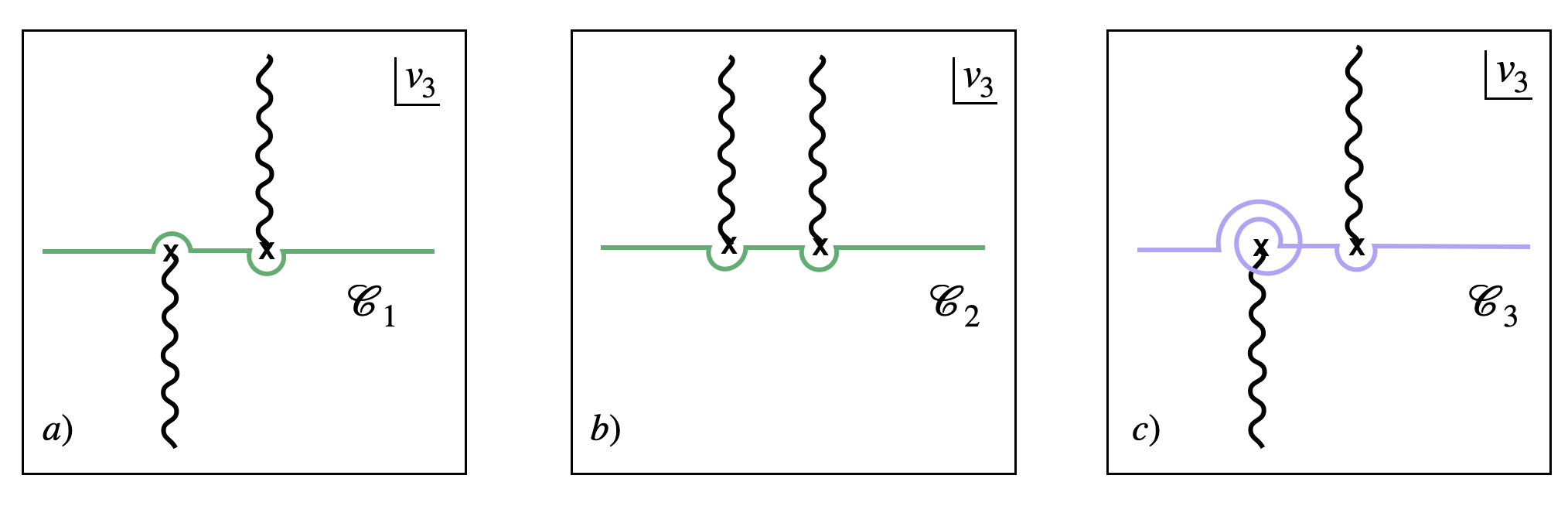}
   \caption{Possible choices of branch cuts in the complex $v_3  \propto \sum_i \omega_i u_i$ plane. Figures $a)$ and $b)$ represent different lifts of the complex plane to its universal cover. Integrating along the real line (green path) corresponds to promoting the integrand in \eqref{eq:branch-int} to respectively $(v_3^2 + v_3 -i\epsilon)$ and $(v_3 - i\epsilon)(v_3 + 1 - i\epsilon)$. Figures a) and c) represent different paths in the same lift of the complex plane to its universal cover . }
   \label{fig:branch-2}
   \end{figure}

Now diagonalizing $A$, we find
\be 
\Lambda_1 = \Lambda_2 = 0, \quad \Lambda_3 = \omega_1^2 + \omega_2^2 + \omega_3^2.
\ee
Normalizing the eigenvectors and defining the associated change of basis matrix $S$, we have
\be 
S B = \left( \eta_3 \frac{s_3 \omega_1 - s_1 \omega_3}{\sqrt{\omega_1^2 + \omega_3^2}}, \eta_2 \frac{s_2 \omega_1 - s_1 \omega_2}{\sqrt{\omega_1^2 + \omega_2^2}} , \eta_3 \frac{s_1 \omega_1 + s_2 \omega_2 + s_3 \omega_3}{\sqrt{\omega_1^2 + \omega_2^2 + \omega_3^2}}\right).
\ee
Furthermore, 
\be 
{\rm det} S = \frac{\omega_1 \Lambda_3}{\sqrt{\omega_1^2 + \omega_3^2}\sqrt{\omega_1^2 + \omega_2^2}\sqrt{\omega_1^2 + \omega_2^2 + \omega_3^2}}.
\ee
Now define $u = S^T v$. Recalling that $\omega_i > 0$, we find
\be 
I = \prod_{i = 1}^3 \frac{1}{\Gamma(\lambda_i)(\eta_i i)^{\lambda_i}} \int_0^{\infty}
 ds_i s_i^{\lambda_i - 1} \int_{-\infty}^{\infty} dv_i  {\rm det}(S) e^{i \Lambda_3 v_3^2 + i \frac{s_1 \omega_1 + s_2 \omega_2 + s_3 \omega_3}{\sqrt{\omega_1^2 + \omega_2^2 + \omega_3^2}} v_3 + i\eta_3 \frac{s_3 \omega_1 - s_1 \omega_3}{\sqrt{\omega_1^2 + \omega_3^2}} v_1 + \eta_2 \frac{s_2 \omega_1 - s_1 \omega_2}{\sqrt{\omega_1^2 + \omega_2^2}} v_2}.
\ee
We can now evaluate the integrals over $v_1$ and $v_2$ to get delta functions, which then are saturated by the $s_3$ and $s_2$ integrals: 
\be 
\begin{split}
I &= (2\pi)^2 \prod_i\left(\frac{1}{\Gamma(\lambda_i)(\eta_i i)^{\lambda_i}} \right)\int_0^{\infty}
 ds_1 s_1^{\lambda_1 - 1} \left(\frac{s_1 \omega_2}{\omega_1}\right)^{\lambda_2 - 1} \left(\frac{s_1 \omega_3}{\omega_1}\right)^{\lambda_3 - 1} \\
 &\times \int_{-\infty}^{\infty} dv_3  \omega_1^{-1} \sqrt{\Lambda_3}  e^{i \Lambda_3 v_3^2 + i \frac{s_1}{\omega_1} \sqrt{\Lambda_3} v_3}.
\end{split}
\ee
At this point we can evaluate the $s_1$ integral  after rescaling $s_1 \rightarrow \omega_1 s_1$ followed by $v_3 \rightarrow s_1 \Lambda_3^{-1/2} v_3$. We see that the $\Lambda_3$ dependence drops out and we obtain
\be
\label{eq:branch-int}
\begin{split}
I = (2\pi)^2 \prod_{i =1}^3 \left( \frac{\omega_i^{\lambda_i - 1}}{\Gamma(\lambda_i)(\eta_i i)^{\lambda_i}}  \right) \frac{1}{2}  \int_{-\infty}^{\infty} dv_3 \frac{i^{\frac{\sum_i \lambda_i - 1}{2}}\Gamma\left(\frac{\sum_i \lambda_i - 1}{2} \right)}{\left(v_3^2 + v_3 \right)^{\frac{\sum_i \lambda_i - 1}{2}}}.
\end{split}
\ee
We now see that the integrand has two branch cuts extending from $0$ and $-1$ to infinity in the complex $v_3$ plane. Note that under our change of variables $v_3 \propto \sum_i \omega_i u_i$. Figure \ref{fig:branch-2} shows different choices of branch cuts and paths in the universal cover of the complex plane. 
 We can now evaluate the integral \eqref{eq:branch-int} for these different paths. For clarity we set
\be
a \equiv \frac{\sum_i \lambda_i - 1}{2}.
\ee
Then 
\be 
I_i = T_i \times 2\pi^2  i^{a}\Gamma\left(a \right) \frac{\Gamma(1-a)^2}{\Gamma(2 - 2a)} \prod_{i =1}^3 \left( \frac{\omega_i^{\lambda_i - 1}}{\Gamma(\lambda_i)(\eta_i i)^{\lambda_i}}  \right),
\ee
where $i$ runs over the three choices of path  shown in Figure \ref{fig:branch-2} and 
\be 
\label{eq:T}
T_1 = -i e^{-i\pi a} \tan\pi a, \qquad T_2 = 0, \qquad T_3 = -2i \sin \pi a = 2 e^{i\pi a} \cos \pi a \times T_1 .
\ee
We now use the Legendre duplication formula to write 
\be 
\Gamma(2 - 2a) = 2^{1 - 2a} \pi^{-1/2} \Gamma(1 - a) \Gamma\left(\frac{3}{2} - a\right),
\ee
together with the Euler identity for the Gamma function to simplify
\be 
I_i = T_i \times i^a 2^{2a}\frac{\pi^{\frac{7}{2}}}{(\sin \pi a)\Gamma\left(\frac{3}{2} - a \right)} \prod_{i =1}^3 \left( \frac{\omega_i^{\lambda_i - 1}}{\Gamma(\lambda_i)(\eta_i i)^{\lambda_i}}  \right). 
\ee

 We would expect path $\mathscr{C}_1$ to yield the scalar celestial amplitude, after appropriate normalization of operators as discussed in Section \ref{sec:preliminaries}. However, what we find instead is 
\be 
 \lim_{R\rightarrow \infty}\prod_{i}\frac{1}{N_{\Delta_i}}\int_{-\infty}^{\infty} du_i (u_i+i\eta_i\epsilon)^{-\lambda_i}\langle{\cal O}_1{\cal O}_2{\cal O}_3\rangle = \frac{- i^{\sum_i \Delta_i - d - 3}}{2 \sin \pi \frac{d + 2 - \sum_i \Delta_i}{2}}\widetilde{A}_3\left(\Delta_1^{\rm CCFT},\Delta_2^{\rm CCFT},\Delta_3^{\rm CCFT}\right).
\ee
Interestingly we can easily deduce from \eqref{eq:T} that lifting the integral to the universal cover of $\mathbb{C}$ and integrating along $\mathscr{C}_3$ in Figure \ref{fig:branch-2} instead leads precisely to the correct normalization, namely
\be 
\lim_{R \rightarrow \infty} \prod_{i}\frac{1}{N_{\Delta_i}}\left. \int_{-\infty}^{\infty} du_i (u_i+i\eta_i\epsilon)^{-\lambda_i}\langle{\cal O}_1{\cal O}_2{\cal O}_3\rangle \right|_{\mathscr{C}_3} = \widetilde{A}_3\left(\Delta_1^{\rm CCFT},\Delta_2^{\rm CCFT},\Delta_3^{\rm CCFT}\right).
\ee
We leave a better understanding of this result to future work.

\bibliographystyle{utphys}
\bibliography{references}

\end{document}